\documentclass[%
 reprint, 
superscriptaddress,
 amsmath,amssymb,
 aps, physrev,
]{revtex4-2}

\usepackage[noend]{algpseudocode}
\usepackage{algorithm}
\usepackage{algorithmicx}
\usepackage{graphicx}   
\usepackage{dcolumn}    
\usepackage{bm}         
\usepackage{amsmath}
\usepackage{theoremref}
\usepackage{enumitem} 
\usepackage{amsthm}
\usepackage{tabularx}
\usepackage{diagbox}
\usepackage{booktabs}
\usepackage{subcaption}
\usepackage{mathtools}
\usepackage{xcolor}
\usepackage{makecell}
\usepackage[export]{adjustbox}
\definecolor{mydarkblue}{rgb}{0,0.08,0.45}

\usepackage[colorlinks=true,linkcolor=mydarkblue,citecolor=mydarkblue,urlcolor=mydarkblue]{hyperref}
\usepackage[capitalise]{cleveref}

\newcommand{\bx}{\mathbf{x}}
\newcommand{\btheta}{\boldsymbol{\theta}}
\newcommand{\bdelta}{\boldsymbol{\delta}}
\newcommand{\bz}{\mathbf{z}}
\newcommand{\rd}{\mathrm{d}}
\newcommand{\bmu}{\boldsymbol{\mu}}
\newcommand{\bSigma}{\boldsymbol{\Sigma}}
\newcommand{\bphi}{\boldsymbol{\varphi}}
\newcommand{\bpsi}{\boldsymbol{\psi}}
\newcommand{\bLambda}{\boldsymbol{\Lambda}}
\newcommand{\bH}{\mathbf{H}}
\newcommand{\bh}{\mathbf{h}}

\usepackage{mdframed}

\newtheorem{theorem}{Theorem}

\begin{document}

\preprint{APS/123-QED}

\title{\textbf{Scalable learning of macroscopic stochastic dynamics} 
}
\author{Mengyi Chen}
\affiliation{Department of Mathematics, National University of Singapore, Singapore, Singapore}
\author{Pengru Huang}
\affiliation{Institute for Functional Intelligent Materials, National University of Singapore, Singapore, Singapore}
\author{Kostya S. Novoselov}
\affiliation{Institute for Functional Intelligent Materials, National University of Singapore, Singapore, Singapore}
\affiliation{ Materials Science and Engineering, National University of Singapore, Singapore, Singapore}
\author{Qianxiao Li}
\email{Contact author: qianxiao@nus.edu.sg}
\affiliation{Department of Mathematics, National University of Singapore, Singapore, Singapore}
\affiliation{Institute for Functional Intelligent Materials, National University of Singapore, Singapore, Singapore}
\date{\today}

\begin{abstract}
Macroscopic dynamical descriptions of complex physical systems
are crucial for understanding and controlling material behavior.
With the growing availability of data and compute, machine learning has 
become a promising alternative to first-principles methods
to build accurate macroscopic models from microscopic trajectory simulations.
However, for spatially extended systems, direct simulations of sufficiently large
microscopic systems that inform macroscopic behavior are prohibitive.
In this work, we propose a framework that learns the macroscopic dynamics of large stochastic microscopic systems using only small-system simulations.
Our framework employs a partial evolution scheme to generate training data pairs by evolving large-system snapshots within local patches. We subsequently derive the closure variables associated with the macroscopic observables and learn the macroscopic dynamics using a custom loss. 
Furthermore, we introduce a hierarchical upsampling scheme that enables efficient generation of large-system snapshots from small-system snapshots.
We empirically demonstrate the accuracy and robustness of our framework through a variety of stochastic spatially extended systems, including those described by stochastic partial differential equations, idealised lattice spin systems,
and a more realistic NbMoTa alloy system.   
\end{abstract}

\maketitle

\section{Introduction}

Macroscopic observables characterize the collective behavior of complex microscopic dynamics and play a crucial role in real-world applications.
They are typically functions of the full microscopic system. 
For example, magnetization is defined as the average of the local magnetic moments, and temperature as the average kinetic energy per degree of freedom.
For alloy systems, macroscopic observables such as thermal conductivity, electrical conductivity, and magnetization are governed by a variety of microscopic interactions, including collective electron scattering mechanisms, lattice vibrations, and microscopic spin coupling interactions~\cite{ordonez2018modeling}. These macroscopic observables capture the material's overall behavior and functionality. 

To obtain accurate time evolution of macroscopic observables,
large-scale microscopic simulations over extended times are often required, yet their high computational expense remains the main bottleneck~\cite{
durrant2011molecular, hollingsworth2018molecular, steinhauser2009review}. 
For example, Density Functional Theory (DFT) and the related \textit{Ab Initio} Molecular Dynamics (AIMD) represent a milestone in computational methods for studying molecules and solid-state materials at the quantum mechanical level. 
However, due to the well-known exponential wall challenge, where computational expense increases rapidly with the number of particles, 
DFT simulations are usually limited to relatively small systems that may be insufficient for capturing true macroscopic behavior~\cite{watson2010rearranging, cohen2012challenges}.
\begin{figure*}[t]
    \centering
    \includegraphics[width=1.0\linewidth]{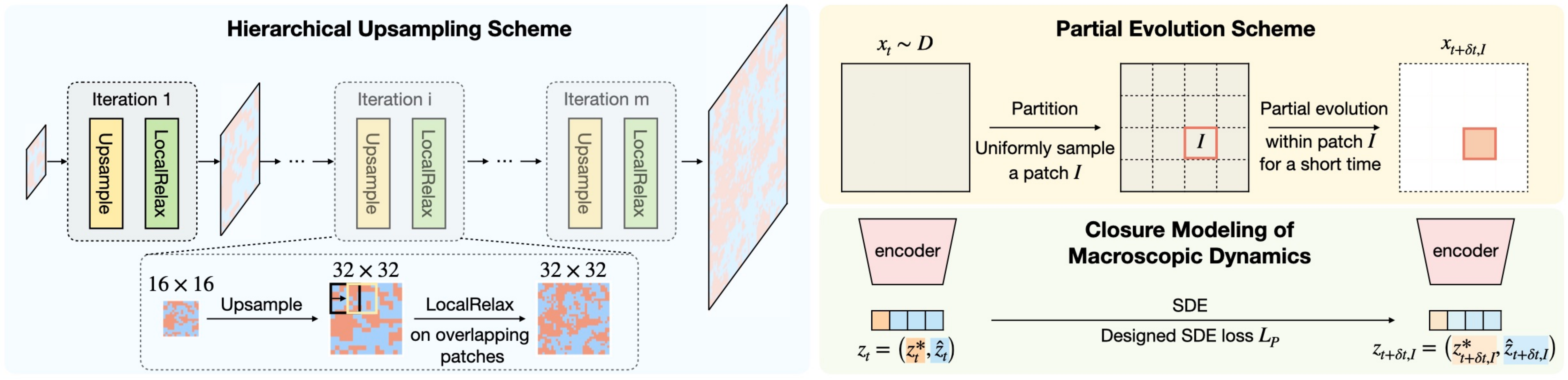}
    \caption{Schematic illustration of our framework.
    The hierarchical upsampling scheme generates the large-system dataset $D$ from the small-system dataset  $D_s$ through multiple iterations, each consisting of an \textsc{Upsample} and a \textsc{LocalRelax} step. An example of one iteration for the Ising model is shown. 
    For the partial evolution scheme, for every $\bx_t \in D$, a patch $\mathcal{I}$ is first uniformly sampled, then the microscopic dynamics is evolved locally within the patch $\mathcal{I}$ for a short time to yield $\bx_{t+\delta t, \mathcal{I}}$.
     For the closure modeling, an autoencoder is trained to discover the closure variables to the macroscopic observables, and the macroscopic dynamics are derived with the designed loss $\mathcal{L}_p$.  }
\label{fig:algorithm}
\end{figure*}

Various approaches have been developed to overcome the computational bottleneck.
The Kinetic Monte Carlo (KMC) algorithm represents the microscopic dynamics as a Markov chain by coarse-graining the time axis~\cite{stamatakis2012unraveling,andersen2019practical,pineda2022kinetic}. The transition rates for all possible events need to be computed for every KMC step. Therefore, KMC is still computationally prohibitive for large systems~\cite{liOpenKMCKMCDesign2019, shangTensorKMCKineticMonte2021}.
Machine learning force fields (MLFFs) replace the expensive \textit{ab initio} force computations with efficient neural network–based predictions~\cite{zhang2018deep, batzner20223, behler2007generalized, gilmer2017neural}.
When applying MLFFs to molecular dynamics simulations, the time step of the molecular dynamics simulation must be chosen on the scale of femtoseconds to capture atomic vibrations. Consequently, millions to billions of integration steps are required for the simulation, and atomic forces are computed for each time step, still resulting in high computational cost for large systems~\cite{jia2020pushing}.
Coarse-grained MLFFs further improve computational efficiency by mapping several atoms onto effective particles, thereby reducing the number of degrees of freedom. 
However, the training of coarse-grained MLFFs requires microscopic simulation data, and the largest barrier to applying coarse-grained MLFFs to large systems is generating enough training data~\cite{durumericMachineLearnedCoarsegrained2023,
husicCoarseGrainingMolecular2020,
kohlerFlowMatchingEfficientCoarseGraining2023,
wangMachineLearningCoarseGrained2019,
zhangDeePCGConstructingCoarsegrained2018}. 
The closure modeling methods, instead, model the dynamics of macroscopic observables directly, but still rely on short microscopic simulations of the large system or microscopic forces on all atoms for macroscopic dynamics derivation of large systems~\cite{chenConstructingCustomThermodynamics2023a}.

Despite their methodological differences, existing approaches all rely either on direct microscopic simulations or on training data derived from such simulations.
However, due to the computational constraint, microscopic simulation of large systems with millions to billions of atoms over extended time is generally intractable. 
This leads to a natural question: Can accurate macroscopic dynamics of large systems be obtained when only small-system microscopic simulations are accessible?
\citet{chen2024learning} proposed a training procedure on the microscopic coordinates
that addresses this question for deterministic dynamics.
However, their method requires partial computation of microscopic forces for the macroscopic dynamics derivation. In the case of stochastic systems, where the dynamics are described by the conditional distribution of the next configuration given the current one, such microscopic forces are generally not well-defined. Hence, their method cannot be easily generalized to stochastic systems.
Yet, stochastic microscopic systems are arguably more prevalent,
especially in the modeling of chemical reactions, molecular dynamics, and ferromagnetic phase transitions~\cite{van1992stochastic}.
The main goal of our method is to address the above question for stochastic dynamical systems. 

In this work, we develop a framework that can accurately derive the macroscopic dynamics of stochastic microscopic systems, while requiring only microscopic simulations of small systems. 
Specifically, given the dataset $D$, composed of snapshots of large systems spanning states from far-from-equilibrium to near-equilibrium, we introduce a partial evolution scheme which evolves $\bx_t \in D$ locally within a local patch $\mathcal{I}$ for a short time $\delta t$ to produce training data pairs $\{\bx_t, \bx_{t+\delta t, \mathcal{I}} \}$. 
Next, building on the workflow of Ref.~\cite{chenConstructingCustomThermodynamics2023a}, we derive the closure variables to the macroscopic observables and model the resulting dynamics via stochastic differential equations (SDE). 
To account for the additional stochasticity introduced by the random selection of patches in the partial evolution scheme, we introduce a modified SDE loss and provide a theoretical justification.
In addition, we design a hierarchical upsampling scheme that efficiently generates a large-system dataset $D$ from a small-system dataset $D_s$, which consists of snapshots sampled from trajectories of the small system.

The key idea of our framework is illustrated in \cref{fig:algorithm}. Key notations are summarized in \cref{app:notation}. We provide a detailed description of each component in \cref{sec:method}. 
In \cref{sec:results}, we further validate the accuracy and robustness of our method through a variety of stochastic microscopic systems, including stochastic partial differential equations (SPDE) systems, spin systems, and a more realistic NbMoTa alloy system.

\section{METHODOLOGY}\label{sec:method}
We focus on microscopic systems that are spatially extended, including SPDE
systems, the Ising model, and alloy systems.
We assume the microscopic time evolution can be modeled as a Markov process of a random variable supported on a finite but large lattice structure. 
Let the microscopic state be $\bx = (\bx_1, \cdots, \bx_n) \in \mathbb{R}^n$,
where $n$ represents the number of
lattice sites of the system. We assume the lattice sites are arranged on a regular
lattice structure, and $\bx_i$ represents some physical quantity associated with
the $i$-th lattice site. To better illustrate this, we provide several examples. In SPDE
systems, $\bx$ can be the state variables after spatial discretization on a
regular grid, with $n$ denoting the total number of grid points. In the Ising
model, $n$ denotes the number of spins, and $\bx_i \in \{-1, 1\}$ represents the
spin state of the $i$-th spin. The spins are arranged on a square lattice. 
In alloy systems, $n$ denotes the number of atoms and $\bx_i$ represents the atom type of the $i$-th lattice site. 
The atoms are arranged on a regular lattice
structure, depending on the crystal structure of the alloy, such as body-centered
cubic (BCC), face-centered cubic (FCC), hexagonal close-packed (HCP), or diamond cubic.

In many real applications, we are interested in the dynamics of some macroscopic observables, denoted by $\bz^{\ast} = \bphi^{\ast}(\bx)$. The form of $\bphi^{\ast}$ is given beforehand, and $\bphi^{\ast}$ can be applied to different system sizes. 
Since the underlying microscopic system is spatially extended, we are interested in the intensive quantities that do not scale with system size. For instance, in the Ising model, it is common to study the average magnetization $M = \sum_{i=1}^n\bx_i/n$ instead of the total magnetization. In what follows, we will limit our discussion of macroscopic observables to intensive quantities.

Assume we are given a microscopic simulator $\mathcal{S}_{n_s}$, which can accurately simulate the microscopic dynamics of a small system up to size $n_s\ll n$ due to computational constraints. 
From this simulator, we obtain the dataset $D_s$ of the small system composed of snapshots sampled from trajectories of the small system.
The goal of this work is to derive the macroscopic dynamics of a large system of size $n$ using only such small-scale simulations. 

\subsection{Closure modeling of macroscopic dynamics}\label{sec:closure modeling}
Existing works on macroscopic dynamics derivation typically involve two components: discovering the closures for macroscopic observables, and jointly deriving their dynamics~\cite{chenConstructingCustomThermodynamics2023a, fu2023simulate, boral2023neural}. 
Our work adopts the same two components for macroscopic dynamics derivation. Assume we are given the dataset $D$ of the large system, consisting of multiple snapshots of the large system, we will first generate temporal training data pairs as follows. 

\paragraph*{Partial evolution scheme.} 
We propose a scheme for locally evolving the microscopic dynamics of a large
system within a small spatial patch, which we refer to as the
\emph{partial evolution scheme}. The purpose of this scheme is to generate locally evolved training data pairs
$\{\bx_t, \bx_{t+\delta t,\mathcal I}\}$ by evolving the microscopic dynamics on
a small patch for a short time interval.

More specifically, we partition the underlying regular lattice into $K = n / n_s$ small patches, each containing $n_s$ lattice sites. For example, for the two-dimensional Ising model on a $64^2$ square lattice, we partition the large square into $64$ square patches of size $n_s = 8^2$.
Let the index set of lattice sites in the $k$-th patch be $\mathcal{I}^k$, which is a subset of $\{1, \cdots, n\}$ and contains $n_s$ lattice sites. The state of the lattice sites in patch $\mathcal{I}^k$ is then written as $\bx_{\mathcal{I}^k} = \{\bx_i\}_{i\in \mathcal{I}^k}$, and the microscopic state as $\bx = (\bx_{\mathcal{I}^1}, \cdots, \bx_{\mathcal{I}^K})$. 
For each configuration $\bx_t \sim D$, we uniformly sample a patch $\mathcal{I}$ from the $K$ patches with probability $1/K$. 
Next, the microscopic simulator $\mathcal{S}_{n_s}$ is used to evolve $\mathbf{x}_{t}$ within the selected patch $\mathcal{I}$ for a short time $\delta t$, yielding the updated state $\bx_{t+\delta t, \mathcal{I}}$. 
Consequently, the resulting training data pair is $\{\bx_t, \bx_{t+\delta  t, \mathcal{I}} \}$.  
We denote the resulting conditional distribution of $\bx_{t+\delta t, \mathcal{I}}$ as $q(\bx_{t+\delta t, \mathcal{I}}|\bx_t)$. Note that the time step $\delta t$ may also be random. For instance, $\delta t$ is sampled from an exponential distribution in the case of kinetic Monte Carlo dynamics. 

The partial evolution scheme is related to patch dynamics in the equation-free
framework (EFF) for multiscale simulation
\cite{multiscakefish,gear2003equation,liu2015acceleration}, in that both approaches
evolve the microscopic dynamics on small spatial patches for short time intervals.
Despite this similarity, we adopt it for a different purpose.
The EFF makes use of patch dynamics as a simulation tool to advance macroscopic observables through repeated microscopic simulations, without explicitly deriving macroscopic dynamical equations. In contrast, we use partial evolution scheme as a data-generation mechanism for learning macroscopic dynamics.

We note that for the partial evolution scheme to be accurate, the microscopic dynamics within a small patch must evolve as if embedded in the full system. 
To achieve this, when evolving the microscopic dynamics within a local patch, appropriate boundary conditions should be imposed.
In our implementation, the local environment is handled in two possible ways. One option is to treat the neighboring sites outside the patch as ghost cells, whose values are held fixed during the short-time evolution. This approach does not increase the computational cost of the partial evolution scheme. 
Alternatively, one may introduce a thin buffer region surrounding the patch. In this approach, the patch and buffer evolve simultaneously, but only the central states are retained for the macroscopic derivation. 
The purpose of the buffer is to shield the interior dynamics from artificial boundary effects. 
This strategy is commonly used in the patch dynamics~\cite{samaeyDampingFactorsGaptooth2003, samaeyGaptoothSchemeHomogenization2003,samaeyPatchDynamicsBuffers2004}. 
While this strategy slightly increases the computational cost, we will keep the buffer thin relative to the full system size, ensuring the cost of the partial evolution scheme remains low.

We will demonstrate how to derive the macroscopic dynamics from the training data pairs obtained from the partial evolution scheme. 

\paragraph*{Autoencoder for discovering closure variables.} 
We employ an autoencoder architecture to discover closure variables $\hat{\bz}$ associated with the macroscopic observables $\bz^{\ast}$. The closure variables will capture the unresolved information by $\bz^{\ast}$, and ensure the dynamics of $\bz =(\bz^{\ast}, \hat{\bz})$ depend only on itself. For example, in our experiments, the dynamics of the Curie-Weiss model can be fully described by the magnetization, hence no closure variables will be needed if $\bz^{\ast}$ represents the magnetization. In contrast, for the Ising model, the magnetization alone cannot capture all the dynamical information, and additional closure variables are required so that the dynamics of $\bz$ only depend on itself.

Since the training data pairs take the form of $\{\bx_t, \allowbreak \bx_{t+\delta t, \mathcal{I}}\}$, 
we want the closure function to be well-defined for both the microscopic state $\bx$ and the microscopic state  $\bx_{\mathcal{I}}$, which are of different dimensions. 
Denote the closure function by $\hat{\bphi}$.
To achieve this, $\hat{\bphi}$ is directly applied to $\bx_{\mathcal{I}}$, yielding $\hat{\bphi}(\bx_{\mathcal{I}})$. The closure representation for the full state $\bx$ is defined as the average of $\hat{\bphi}(\bx_{\mathcal{I}})$ over all the patches:
\begin{equation}
    \textstyle
    \hat{\bphi}(\bx) = \frac{1}{K} \sum_{\mathcal{I}\in \{\mathcal{I}^1, \cdots, \mathcal{I}^K\}} \hat{\bphi}(\bx_{ \mathcal{I}}).
\end{equation}
We denote the closure variables by $\hat{\bz} = \hat{\bphi}(\bx)$, and concatenate it with the macroscopic observable $\bz^{\ast} = \bphi^{\ast}(\bx)$ to form the full latent state $\bz = (\bz^{\ast}, \hat{\bz})= (\bphi^{\ast}(\bx), \hat{\bphi}(\bx))$.
By definition of the closure variables, $\hat{\bz}$ are intensive quantities, just like the macroscopic observables $\bz^{\ast}$. Therefore, $\bz$ represents intensive quantities. 
Denote the encoder by $\bphi = (\bphi^{\ast}, \hat{\bphi})$ and the decoder by $\bpsi$, where $\bphi^{\ast}$ is the predefined macroscopic observable function with no trainable parameters. The functions $\hat{\bphi}$ and $\bpsi$ are parameterized by neural networks and are trained jointly. We omit the explicit dependence on the parameters for notational simplicity.
The autoencoder is trained by minimizing the reconstruction loss:
\begin{equation}
    \mathcal{L}_{\text{recon}} = \mathbb{E}_{\bx_t}|| \bpsi\circ \bphi(\bx_t)- \bx_t  ||_2^2 .
\end{equation}
Once the autoencoder is trained, we will generate the latent training data pairs $\{\bz_t, \bz_{t+\delta t, \mathcal{I}} \}$ for the macroscopic dynamics derivation:
\begin{equation}
    \label{eq:latent}
    \begin{aligned}
        &\bz_t = \bphi(\bx_t),  \\
        &\bz_{t+\delta t, \mathcal{I}} \coloneqq \bz_t + \left(\bphi(\bx_{t+\delta t, \mathcal{I}}) - \bphi(\bx_{t, \mathcal{I}}) \right),
    \end{aligned}
\end{equation}
where $\bx_{t, \mathcal{I}}$ denotes the restriction of $\bx_t$ to the local patch $\mathcal{I}$. 
Next, we introduce the process of deriving macroscopic dynamics. 

\paragraph*{Macroscopic dynamics derivation. }
We model the macroscopic dynamics with SDE:
\begin{equation}\label{eq:SDE}
    \mathrm{d}\bz_t = \bmu(\bz_t)\rd t + \bSigma^{1/2}(\bz_t) \rd \mathbf{B}_t,
\end{equation}
where $\bmu$ is the drift term and $\bSigma$ is the diffusion term.
In most experiments, we adopt fully connected networks for both $\bmu$ and $\bSigma$. 
For the NbMoTa alloy experiment in \cref{exp:alloy}, we adopt OnsagerNet~\cite{yu2021onsagernet, chenConstructingCustomThermodynamics2023a} for $\bmu$ to account for more complex macroscopic dynamics.
The OnsagerNet can capture physically interpretable and stable macroscopic dynamics by incorporating the generalized Onsager principle into the model structure, and its form is given by: 
\begin{equation}
    \bmu(\bz_t) = -(\mathbf{M}(\bz_t) + \mathbf{W}(\bz_t))\nabla V(\bz_t) + \mathbf{f}(\bz_t),
\end{equation}
where $\mathbf{M}$ is a symmetric positive semi-definite matrix describing energy dissipation,  $\mathbf{W}$ is a skew-symmetric matrix describing energy conservation, $V$ is a potential function, and $\mathbf{f}$ is a vector field representing external forces. 

Existing works train the SDE by minimizing the negative log-likelihood~\cite{chenConstructingCustomThermodynamics2023a, dietrichLearningEffectiveStochastic2023, zhang2022gfinns, gaoLearningInterpretableDynamics2024}:
\begin{equation}\label{eq:L}
\begin{aligned}
    \mathcal{L}[\bmu, \bSigma] &= \mathbb{E}_{\bz_t, \bz_{t+\delta t}}
    [ - 2\log \\
    &\quad p \left( \bz_{t+\delta t} \,\big|\, \bz_t + \bmu(\bz_t) \delta t, \, \bSigma(\bz_t) \delta t \right) ],
\end{aligned}
\end{equation}
where $\bz_{t+\delta t}$ denotes the latent state obtained by evolving the full system from $\bz_t$ over a time step $\delta t$.
The conditional distribution $p$ is given by the Gaussian distribution $\mathcal{N}(\bz_{t+\delta t} ; \bz_t + \bmu(\bz_t)\delta t, \bSigma(\bz_t)\delta t)$, obtained by discretizing the SDE with the Euler-Maruyama scheme.

In our setting, however, $\bz_{t+\delta t, \mathcal{I}}$ is not obtained by evolving the full system by $\delta t$. 
Instead, it results from a partial evolution over a localized spatial patch. 
To account for this, we adapt the SDE loss as follows:
\begin{equation}\label{eq:L_p}
    \begin{aligned}
        \mathcal{L}_p[\bmu, \bSigma] &= \mathbb{E}_{\bz_t, \bz_{t+\delta t,\mathcal{I}}}
    [-2\log \\
    &\quad p ( \bz_{t+\delta t, \mathcal{I}} \,|\, \bz_t + \bmu(\bz_t)\delta t,\, K \bSigma(\bz_t)\delta t ) ] .
    \end{aligned}
\end{equation}

The only difference between $\mathcal{L}$ and $\mathcal{L}_p$ is that the covariance term in $\mathcal{L}_p$ is multiplied by a factor $K$, where $K$ denotes the number of patches introduced earlier.
We can interpret the influence of $K$ qualitatively. 
The factor $K$ in the loss compensates for the additional stochasticity, leading to a smaller learned diffusion term. 
During the data generation of $\bx_{t+\delta t, \mathcal{I}}$ from $\bx_t$, we introduce additional randomness by performing partial evolution. Therefore, in the derivation of macroscopic dynamics, we multiply the diffusion term by $K$ to correct for the extra stochasticity. 

To learn the macroscopic dynamics, we parametrize both $\bmu$ and $\bSigma$ with neural networks, denoted by $\bmu_{\btheta}$ and $\bSigma_{\btheta}$, respectively. 
The training objective is given by the loss function:
\begin{equation}
\mathcal{L}_p(\btheta) \coloneqq \mathcal{L}_p[\bmu_{\btheta}, \bSigma_{\btheta}].
\end{equation}
We minimize $\mathcal{L}_p(\btheta)$ to obtain the optimal parameter.

 It is important to observe that in \cref{eq:latent}, we define $\bz_{t+\delta t,\mathcal{I}}$ in a nontrivial way, while the most naive way will be $\bz_{t+\delta t,\mathcal{I}}=\bphi(\bx_{t+\delta t, \mathcal{I}})$. Note that the loss function $\mathcal{L}_p(\btheta)$ involves the weighted norm of the residual $\bz_{t+\delta t, \mathcal{I}} - \bz_t -\bmu(\bz_t)\delta t$ since $p$ represents a Gaussian distribution. 
 If we define $\bz_{t+\delta t,\mathcal{I}}=\bphi(\bx_{t+\delta t, \mathcal{I}})$ in the naive way, we will have: 
\begin{equation}\label{eq:naive}
 \textstyle
     \bz_{t+\delta t, \mathcal{I}} - \bz_t = \frac{1}{K}\sum_{\mathcal{J}} (\bphi(\bx_{t+\delta t, \mathcal{I}})- \bphi(\bx_{t, \mathcal{J}} )),
 \end{equation}
 which is the average of $\bphi(\bx_{t+\delta t, \mathcal{I}})- \bphi(\bx_{t, \mathcal{J}} )$. Since $\bphi(\bx_{t+\delta t, \mathcal{I}})$ may be very different from $\bphi(\bx_{t, \mathcal{J}} )$ when $\mathcal{J} \neq \mathcal{I}$, the resulting $\bz_{t+\delta t, \mathcal{I}} - \bz_t$ will be very noisy.  
 However, the definition in \cref{eq:latent} will yield:
\begin{equation}\label{eq:nontrivial}
     \bz_{t+\delta t, \mathcal{I}} - \bz_t = \bphi(\bx_{t+\delta t, \mathcal{I}})- \bphi(\bx_{t, \mathcal{I}}),
 \end{equation}
which directly measures the change in the patch $\mathcal{I}$, thus reducing noise and leading to more stable training.  
We refer to the approach that adopts the naive formulation $\bz_{t+\delta t,\mathcal{I}} = \bphi(\bx_{t+\delta t, \mathcal{I}})$ and derives the macroscopic dynamics via $\mathcal{L}$ as the baseline. 
A detailed comparison of our method with the baseline is provided in \cref{sec:results}.

\paragraph*{Theoretical justification.} 
In many microscopic systems, the microscopic interactions between lattice sites are local. For instance, microscopic interactions are limited to the first nearest neighbor in the Ising model, and are limited to a finite cutoff distance in alloy systems. 
Under this locality assumption, when the time increment $\delta t$ is sufficiently small,
the state increments on disjoint spatial patches are approximately independent.
Specifically, define the short-time increments
\begin{equation}
\begin{aligned}
&\Delta \bz_t 
\coloneqq \bz_{t+\delta t} - \bz_t,
&\,
&\Delta \bz_t^{\ast} 
\coloneqq \bz^{\ast}_{t+\delta t} - \bz^{\ast}_{t}, \\
&\Delta \bz_{t,\mathcal{I}} 
\coloneqq \bz_{t+\delta t,\mathcal{I}} - \bz_{t},
&\,
&\Delta \bz^{\ast}_{t,\mathcal{I}} 
\coloneqq \bz^{\ast}_{t+\delta t,\mathcal{I}} - \bz^{\ast}_{t} .
\end{aligned}
\end{equation}
Then, for two disjoint patches $\mathcal{I} \neq \mathcal{J}$, we have
\begin{equation}
    \begin{aligned}
    \Delta \bz_{t,\mathcal{I}} &= \bphi(\bx_{t+\delta t,\mathcal{I}}) - \bphi(\bx_{t,\mathcal{I}}), \\
    \Delta \bz_{t,\mathcal{J}} &= \bphi(\bx_{t+\delta t,\mathcal{J}}) - \bphi(\bx_{t,\mathcal{J}}),
    \end{aligned}
\end{equation}
which can be treated as approximately independent random variables.

Let $\hat{q}(\bx_{t+\delta t, \mathcal{I}}|\bx_t)$ denote the distribution of $\bx_{t+\delta t, \mathcal{I}}$ derived by evolving the full system from $\bx_t$ for a time step $\delta t$ and subsequently restricting the state to the local patch $\mathcal{I}$.
Under the local interaction assumption and for sufficiently small $\delta t$, the distribution $q$ can be well approximated by the partial evolution distribution $\hat{q}$: 
\begin{align}\label{eq:equal_distribution}
q(\bx_{t+\delta t, \mathcal{I}}|\bx_t) \approx \hat{q}(\bx_{t+\delta t, \mathcal{I}}|\bx_t).
\end{align}
Furthermore, since we consider a sufficiently large microscopic system and the macroscopic observables are intensive quantities, 
the macroscopic observable of the full system can be approximated by the average over local patches:
\begin{equation}\label{eq:mean_macro_observable}
\textstyle
\bphi^{\ast}(\bx) \approx \frac{1}{K} \sum_{\mathcal{I}} \bphi^{\ast}(\bx_{\mathcal{I}}).
\end{equation}
Similarly, the macroscopic increment
$
    \Delta \bz^{\ast}_{t}
    = \bphi^{\ast}(\bx_{t+\delta t}) - \bphi^{\ast}(\bx_{t})
$
can be approximated by the average of the local increments,
\begin{equation}\label{eq:mean_macro_increment}
\textstyle
\Delta \bz^{\ast}_{t}
\approx \frac{1}{K}\sum_{\mathcal{I}} \Delta \bz^{\ast}_{t,\mathcal{I}} .
\end{equation}
\cref{eq:mean_macro_observable,eq:mean_macro_increment} hold exactly when the macroscopic observable can be written as the mean of a function depending only on individual lattice sites, such as the magnetization.
For more complex macroscopic observables, the equality is only approximate, but the approximation improves and becomes accurate in the limit of a large system size.

We provide a theoretical justification for the loss $\mathcal{L}_p$ under the above conditions:
\begin{theorem}\label{thm:1}
Assume that for any two disjoint patches $\mathcal{I} \neq \mathcal{J}$, the short-time increments
$\Delta \bz_{t,\mathcal{I}}$ and $\Delta \bz_{t,\mathcal{J}}$ are conditionally independent given $\bx_t$.
Assume further that the encoder $\boldsymbol{\varphi}$ is uniformly bounded, i.e.,
$\|\boldsymbol{\varphi}\|_{\infty} \le M$ for some $M > 0$.

Suppose that \cref{eq:equal_distribution,eq:mean_macro_increment} hold up to second-order accuracy in $\delta t$,
in the sense that the total variation distance between $q$ and $\hat{q}$ satisfies
\begin{equation}
\delta_{\mathrm{TV}}\!\left(q(\cdot \mid \bx_t), \hat{q}(\cdot \mid \bx_t)\right)
\le C_1 \delta t^2,
\end{equation}
and that the macroscopic increment approximation error satisfies
\begin{equation}
\textstyle
\bigl\| \mathbb{E}_{\bx_{t+\delta t}\mid \bx_t} [
\Delta \bz_t^{\ast}
- \tfrac{1}{K}\sum_{\mathcal{I}} \Delta \bz^{\ast}_{t,\mathcal{I}}]
\bigr\|
\le C_2 \delta t^2 ,
\end{equation}
where $C_1, C_2 > 0$ are constants and
$\bx_{t+\delta t}\mid \bx_t$ denotes the ground-truth conditional distribution
obtained by evolving the full system from $\bx_t$ for time $\delta t$.
 
Under these assumptions, the functional $\mathcal{L}[\bmu,\bSigma]$ admits a unique minimizer
$(\bmu^{\ast}, \bSigma^{\ast})$, and the functional $\mathcal{L}_{p}[\bmu,\bSigma]$
admits a unique minimizer $(\bmu^{\dagger}, \bSigma^{\dagger})$, such that
\begin{equation}
\begin{aligned}
\|\bmu^{\ast} - \bmu^{\dagger}\|_{\infty} &\le K_1 \delta t, \\
\|\bSigma^{\ast} - \bSigma^{\dagger}\|_{\infty} &\le K_2 \delta t ,
\end{aligned}
\end{equation}
where $K_1, K_2 > 0$ are constants depending only on $C_1$, $C_2$, and $M$.
In particular, if $q = \hat{q}$ holds exactly, we have
\begin{equation}
\bmu^{\ast} = \bmu^{\dagger},
\qquad
\bSigma^{\ast} = \bSigma^{\dagger}.
\end{equation}
\end{theorem}

We emphasize that the independence assumption in \cref{thm:1} is about short-time increments on disjoint patches conditioned on $\bx_t$, not about independence of the patch states themselves.
The proof is based on a direct computation of the first and second variations of the loss functions. We provide the full proof in \cref{app:a}. \cref{thm:1} theoretically justifies our framework:
Under appropriate conditions, the macroscopic dynamics learned from data generated via partial evolution are as accurate as those learned from full, computationally expensive microscopic simulations. 
In particular, for the stochastic Predator–Prey system in \cref{exp:PP}, the assumptions of \cref{thm:1} are satisfied, as verified in \cref{app:proof}.

\subsection{Hierarchical upsampling scheme}

\begin{figure}[!htbp]
\begin{algorithm}[H]
    \caption{Hierarchical Upsampling Scheme}
    \label{alg:upsampling}
    \begin{algorithmic}[1]
        \Require{ \Statex $D_s$: dataset of small-system snapshots
        \Statex $N_{\text{iter}}$: number of iterations}
        \State Initialize $D^{(0)} \gets D_s$
        \For{$i = 1$ {\bfseries to} $N_{\text{iter}}$}
            \State $D^{(i)} \gets \textsc{LocalRelax} \big(\textsc{Upsample}(D^{(i-1)})\big)$
        \EndFor
        \State $D \gets D^{(N_{\text{iter}})}$
        \State \Return $D$ 
        \Comment {dataset of large-system snapshots}
    \end{algorithmic}
\end{algorithm}
\vspace{-20pt}
\end{figure}
In \cref{sec:closure modeling}, we assume access to the dataset $D$, which contains multiple snapshots of large systems. In practice, however, direct access to $D$ is typically unavailable, as only microscopic simulations of small systems can be performed. To address this, 
we introduce a hierarchical upsampling scheme for generating the large-system dataset $D$ from the small-system dataset $D_s$. 
The hierarchical upsampling scheme is illustrated in \cref{alg:upsampling}. It consists of multiple iterations, each involving two steps: \textsc{Upsample} and \textsc{LocalRelax}. In the \textsc{Upsample} step, the configurations in $D^{(i)}$ are expanded into configurations of size $m$ times larger, where $m \geq 2$ is an integer. Next, we apply a \textsc{LocalRelax} step to remove the unphysical artifacts that are introduced in the \textsc{Upsample} step. More specifically, each generated large-system configuration is divided into overlapping patches of size $n_s$, and short-time relaxation or local dynamics evolution is applied within each patch. 

For example, consider the two-dimensional Ising Model to be introduced in \cref{exp:ising}, where the microscopic dynamics is chosen to be the continuous-time Glauber dynamics. 
We assume direct simulations are feasible only for systems up to size $n_s = 8^2$. Starting from the small-system dataset $D_s$, we apply the hierarchical upsampling scheme for $N_{\text{iter}}=3$ iterations to obtain the large-system dataset $D$ of size $n = 64^2$. 
In the first iteration, the \textsc{upsample} step replicates each spin into a $m=2^2$ block, yielding a $16^2$ dimensional system. Next, we apply the \textsc{LocalRelax} step by dividing each $16^2$ dimensional configuration into $16$ patches of size $n_s=8^2$ with a stride of $4$. 
Within each patch, we run the continuous-time Glauber dynamics for a short time to remove the unphysical artifacts introduced in the \textsc{upsample} step. 
In the second iteration, starting from the $16^2$ dimensional dataset $D^{(1)}$, we repeat the \textsc{upsample} and \textsc{LocalRelax} step to obtain $D^{(2)}$ for the $32^2$ dimensional system. 
Subsequently, in the third iteration, we obtain the target large-system dataset $D$ of size $64^2$ from $D^{(2)}$.
We provide a graphical illustration of one iteration for the Ising model in \cref{fig:algorithm} (a).
The concrete form of \textsc{Upsample} and \textsc{LocalRelax} may vary for different microscopic systems, and further details are given in \cref{sec:results}. 

By generating training data through a hierarchical upsampling and partial evolution scheme, our method circumvents the expensive, large-system microscopic simulations that are required by most of the existing methods. The main computational savings of our method come from the efficient generation of training data.

\section{RESULTS}\label{sec:results}
In this section, we empirically validate the accuracy and robustness of our method across various microscopic systems. We first demonstrate our method on a SPDE system and spin systems, and then validate it on a more realistic NbMoTa alloy system. 

\subsection{Stochastic Predator-Prey
system\label{exp:PP}}
\begin{figure}[t]
    \centering
    \includegraphics[width=1.0\linewidth]{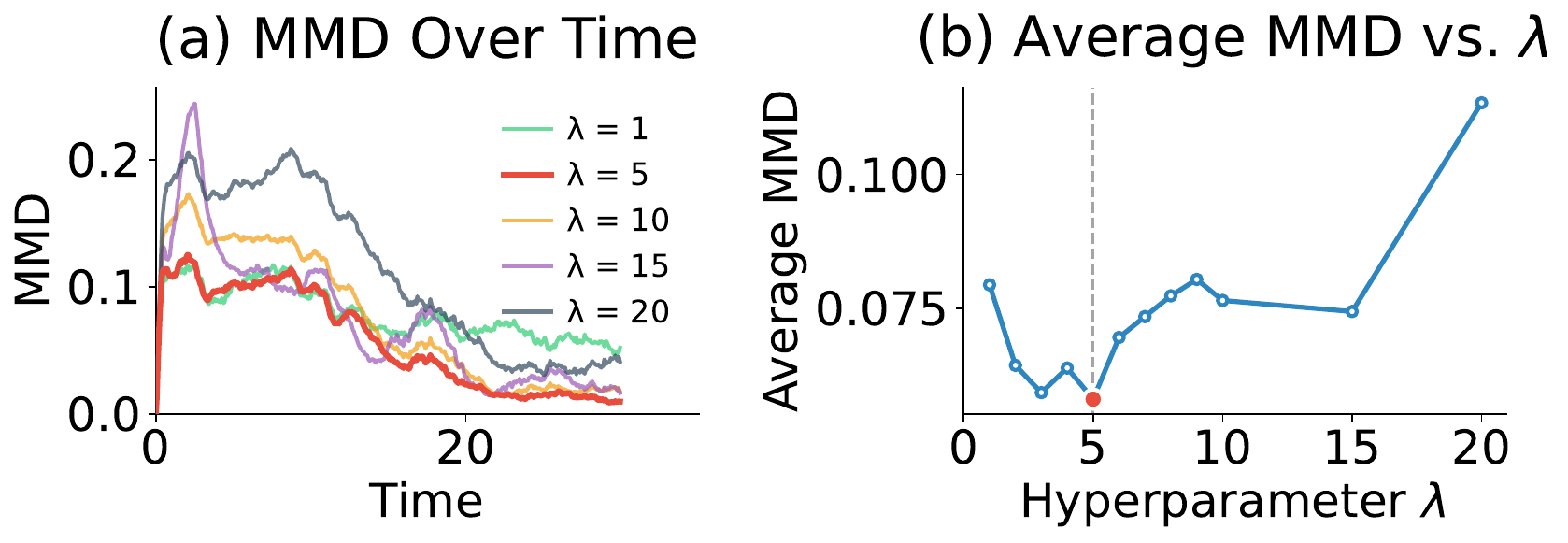}
    \caption{Results on the stochastic Predator-Prey system. (a) The MMD is plotted as a function of time. (b) The average MMD over the entire simulation time is reported as a function of the hyperparameter $\lambda$. }
    \label{fig:PP}
\end{figure}
We first consider a one-dimensional SPDE system, mainly to validate the correctness of our method and investigate the impact of the coefficient $K$ in the designed loss $\mathcal{L}_p$. 
The stochastic Predator-Prey system is given by \cref{eq:PP}, where $u, v$ denote the dimensionless populations of the prey and predator, and $\xi_u, \xi_v $ are independent space-time white noise terms.  In our experiment,  we set the parameters $a=3, b=0.4, c=0$ and $\sigma_u = \sigma_v = 0.02$. 
\begin{equation}\label{eq:PP}
\begin{aligned}
&\frac{\partial u}{\partial t} = u(1-u-v) + c \frac{\partial^2 u}{\partial x^2} 
+ \sigma_u \,\xi_u(t,x), \\
&\frac{\partial v}{\partial t} = a v(u-b) + \frac{\partial^2 v}{\partial x^2} 
+ \sigma_v \,\xi_v(t,x),  \\
&x \in \Omega = [0,1], \; t \ge 0,
\end{aligned}
\end{equation}
We impose Neumann boundary conditions: 
\begin{equation}
    \frac{\partial u}{\partial x}(t,0) = \frac{\partial u}{\partial x}(t,1) = 0, \frac{\partial v}{\partial x}(t,0) = \frac{\partial v}{\partial x}(t,1) = 0.
\end{equation}
The initial conditions are defined as:
\begin{equation}
\begin{aligned}
    u(x, 0) &= c_1 + c_2 \cos(10 \pi x) ,\\
    v(x, 0) &= c_1 - c_2\cos(10\pi x ) ,
\end{aligned}
\end{equation} 
where $ c_1 \sim \mathcal{U}(0.05, 0.15), c_2 \sim \mathcal{U}(0.45, 0.55)$. 

For training data generation, we first generate the small-system dataset $D_s$ by discretizing the spatial domain $\Omega$ into $100$ uniform grids. We then solve \cref{eq:PP} from $t=0$ to $T=30$ with time step $\delta t=0.01$.
The small system thus contains $n_s =100$ lattice sites. 
The goal is to learn the macroscopic observables of a large system discretized on $n=200$ uniform grid points. 
To construct the large-system dataset $D$, we perform an \textsc{Upsample} step by linearly interpolating the solution from the coarse grid with $100$  points onto a finer grid with $200$ points. 
Compared to snapshots sampled from large-system trajectories, snapshots generated by linear interpolation are smoother. However, we expect the resulting differences to be small. Therefore, for simplicity, we do not apply the \textsc{LocalRelax} step.

The $n=200$ uniform grids are partitioned into $K=5$ patches, each containing $40$ grids. For each $\bx_t \sim D$, we first uniformly sample a patch with $p=1/5$, and subsequently evolve the system locally within the patch for one time step $\delta t = 0.01$ to obtain the updated state $\bx_{t+\delta t, \mathcal{I}}$.  

The macroscopic observable $\bz^{\ast}$ of interest is chosen to be the mean of the populations of the prey and predator over the spatial grid points, which is two-dimensional. We derive another 2 closure variables using an autoencoder, hence the dimension of $\bz$ is 4. The only difference between $\mathcal{L}$ and $\mathcal{L}_p$ lies in the coefficient multiplying the diffusion term. We treat the coefficient as a hyperparameter $\lambda$ and train the SDE with different values of $\lambda$. Theoretical analysis in \cref{thm:1} shows that the optimal value of $\lambda$ is $K$ under appropriate conditions. We explore various values of $\lambda$ to empirically investigate the influence of $\lambda$. 

\cref{fig:PP} shows the result on the $9$ test datasets with different combinations of parameters $(c_1, c_2) \in \{0.05, \allowbreak 0.10, 0.15\} \times \{0.45, 0.50, 0.55\}$. Each test dataset consists of $50$ trajectories with the same initial condition. 
Once the SDE model in \cref{eq:SDE} is trained, we simulate it for a long time with the Euler-Maruyama method starting from the same initial condition as the test dataset. 
We employ Maximum Mean Discrepancy (MMD)~\cite{gretton2012kernel, kidger2021neural} to quantify the discrepancy between the predicted and ground-truth trajectories, which is widely used for comparing probability distributions. We use a mixture of radial basis function (RBF) kernels with varying bandwidths to improve the robustness of MMD.
For each time point, we calculate the MMD between the marginal distributions of the predicted trajectories and the ground truth trajectories, and report the results averaged over all $9$ test datasets.   

From \cref{fig:PP} we can observe that when $\lambda=K=5$, the predicted trajectories achieve the minimal discrepancy from the ground truth trajectories, which correspond well with our theoretical analysis. In fact, for the stochastic Predator-Prey system, the assumptions in \cref{thm:1} are exactly satisfied. Therefore, it is expected that the optimal hyperparameter $\lambda$ is equal to $K$. In subsequent experiments with more complex microscopic dynamics, the assumptions in \cref{thm:1} may only hold approximately. Thus, the optimal value of $\lambda$ may deviate slightly from $K$. We will treat $\lambda$ as a tunable hyperparameter and perform a hyperparameter search initialized from $K$ in the following experiments.

\subsection{Ising model}\label{exp:ising}
\begin{figure}[t]
    \centering
    \includegraphics[width=0.6\linewidth]{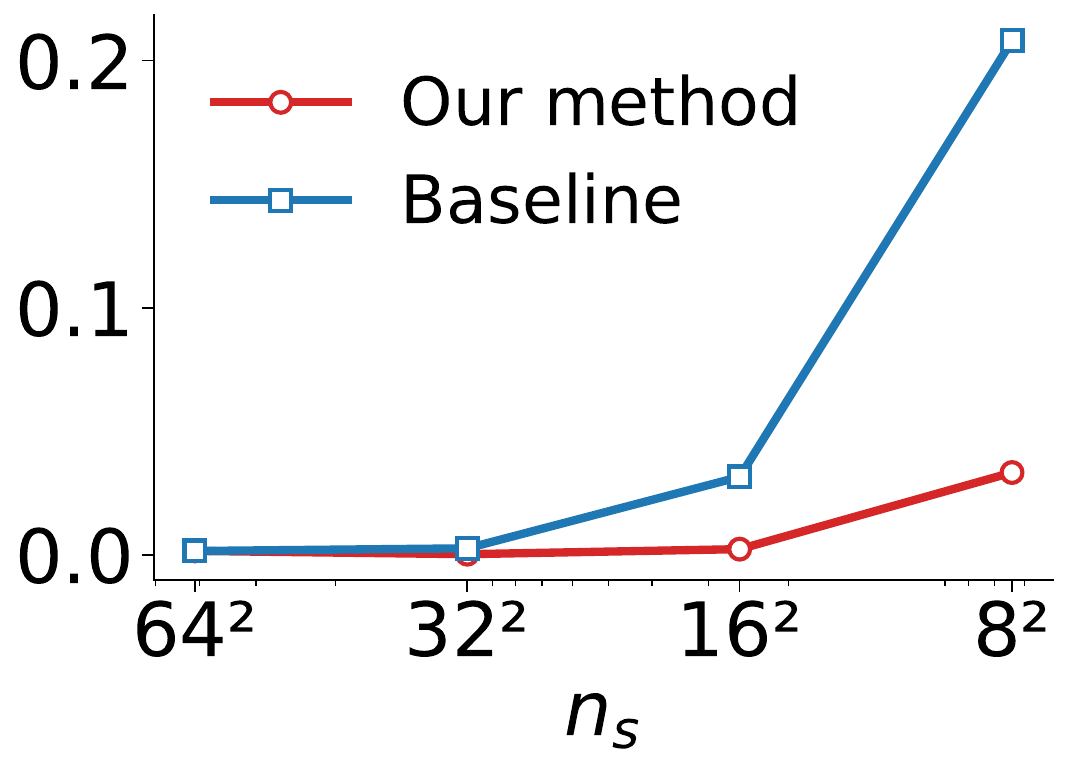}
    \caption{Results on the Ising model, where the test error is plotted as a function of $n_s$. The test error is the mean relative error of the mean macroscopic observables between ground-truth and predicted trajectories.}
    \label{fig:Ising_scale}
\end{figure}
\begin{figure}[t]
    \centering
    \includegraphics[width=1.0\linewidth]{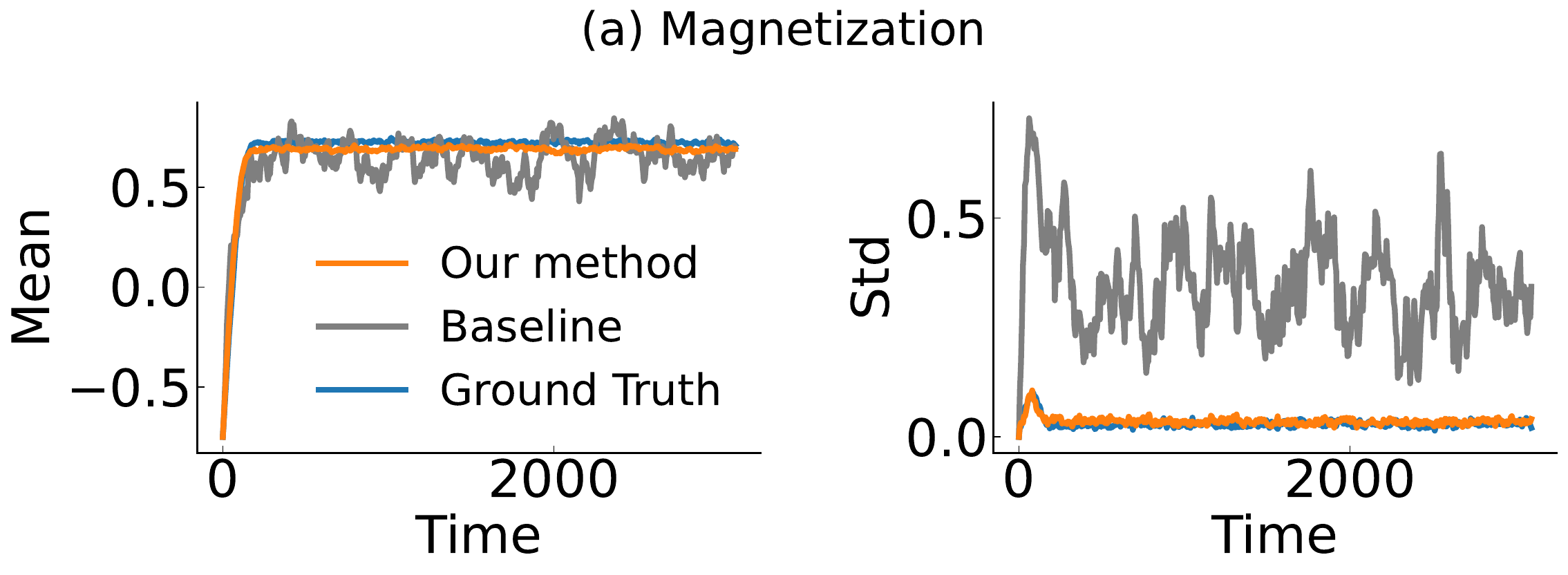}\\
    \includegraphics[width=1.0\linewidth]{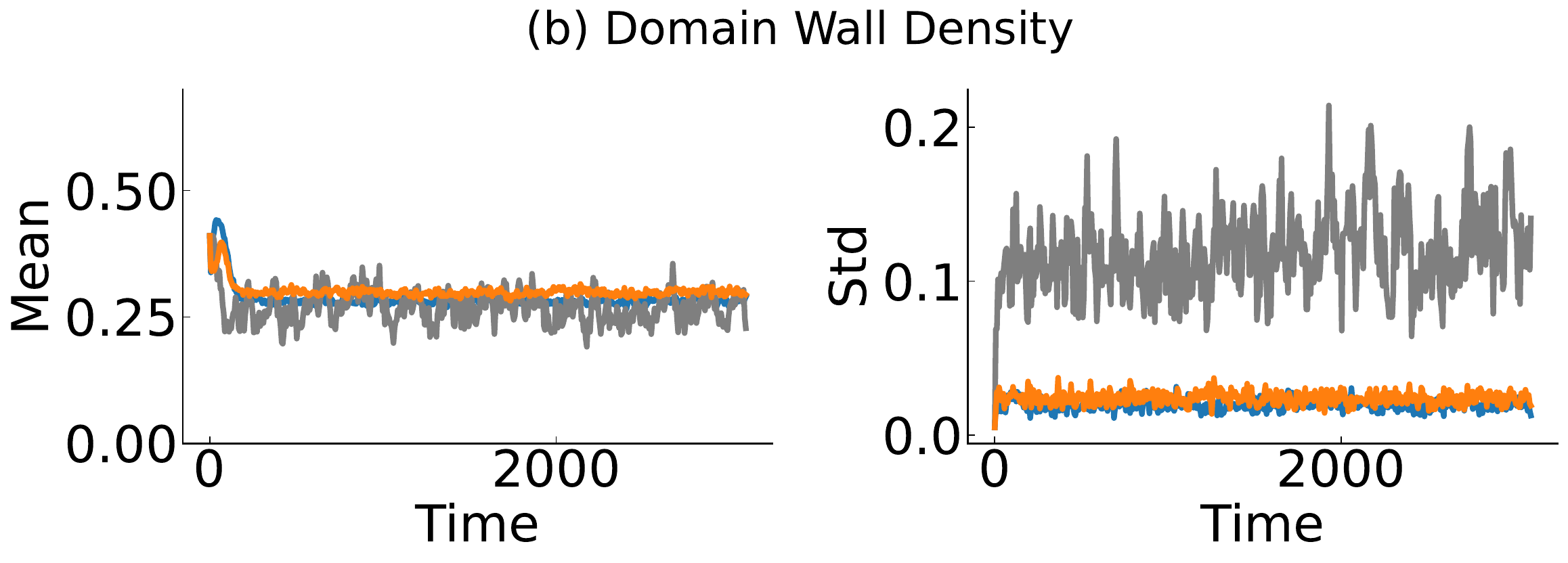}
    \caption{Results on the Ising model with $n_s=16^2$. 
    Mean and standard deviation are estimated from $20$ trajectories per method. 
    (a) Magnetization statistics. (b) Domain wall density statistics. }
    \label{fig:ising_traj}
\end{figure}

\begin{figure*}[!t]
    \centering
    \includegraphics[width=1.0\linewidth]{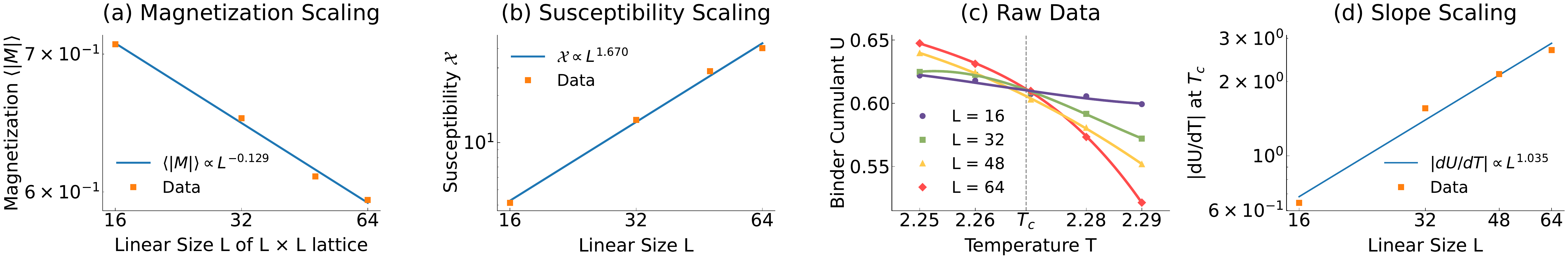}
    \caption{Finite size scaling results of the Ising model at $T=2.27$, which is very close to $T_c\approx 2.269$. (a) Scaling of the equilibrium magnetization, with a fitted value $(\beta/\nu)^{\ast} = 0.129$. (b) Scaling of the magnetic susceptibility, with a fitted value $(\gamma/\nu)^{\ast} = 1.670$. (c) Raw Binder cumulant data for different temperatures and linear size $L$. Linear fitting is used for $L=16$, and cubic polynomial fitting for  $L=32, 48, 64$. The slope at $T_c$ is extracted from the fitted curve.  (d) Log-log plot of $|dU/dT|_{T=T_c}$ versus $L$, with a fitted value $\nu^{\ast} = 1/1.035 \approx 0.97$. }
    \label{fig:ising_scaling}
\end{figure*}

\begin{figure*}[!t]
    \centering
    \includegraphics[width=1.0\linewidth]{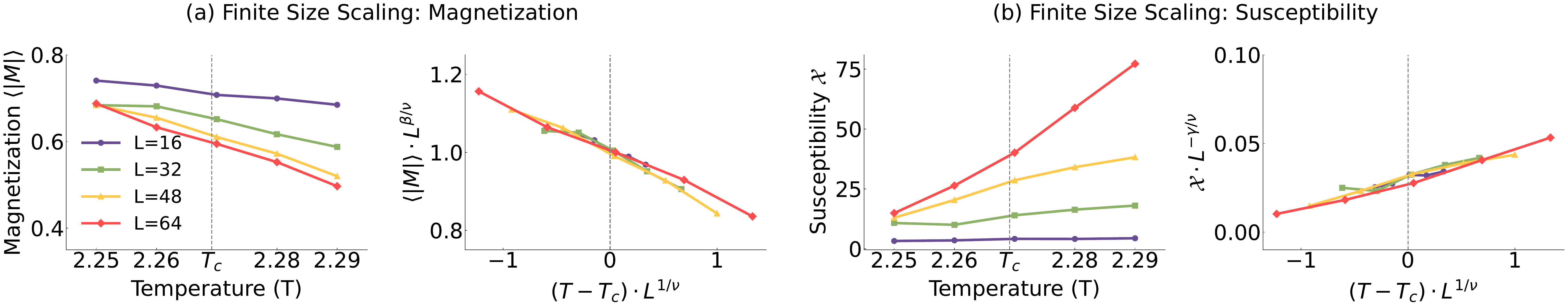}
    \caption{Finite size scaling results of the Ising model across various temperatures and system sizes. The data are collapsed using the theoretical critical exponents. 
    (a) Finite size scaling analysis of the magnetization data calculated from the predicted trajectories by our method. (b) Finite size scaling analysis of the susceptibility. }
    \label{fig:finite_size_scaling}
\end{figure*}
Having validated our method on a toy SPDE system, we now turn to more complex spin systems. We first consider the Ising model, which plays an important role in statistical physics for investigating order-disorder phase transitions and critical phenomena~\cite{brush1967history, cipra1987introduction}. 
The experiments are divided into two parts. 
In the first part, we conduct an ablation study on the computational power $n_s$ of the microscopic simulator $\mathcal{S}_{n_s}$, and compare the performance of our method with the baseline across different values of $n_s$. 
In the second part, we further evaluate the ability of our method to accurately capture the critical behavior of macroscopic dynamics. 

We consider the two-dimensional Ising model, where the spins are arranged on a square lattice of size $n = L \times L$. 
The Hamiltonian of the Ising model is given by:
\begin{equation}
\label{eq:hamiltonian_ising}
    H_{n, h}(\sigma) = -\frac{J}{2} \sum_{\langle i,j\rangle} \sigma_i \sigma_j - h \sum_{i=1}^n \sigma_i ,
\end{equation}
where $J>0$ denotes the interaction strength, $h$ denotes the external magnetic field, $\langle i,j\rangle$ represents nearest-neighbor pairs, and $\sigma_i \in \{-1, 1\}$ denotes the spin at site $i$. 
The microscopic state is denoted by $\bx = \{\sigma_1, \cdots, \sigma_n\}$.
Throughout this paper, we employ dimensionless units by setting $J = 1$ and the Boltzmann constant $k_B=1$. 
We describe the microscopic evolution of the system using continuous-time Glauber dynamics, as detailed in \cref{app:b}.
The critical temperature of the two-dimensional Ising model is $T_c\approx 2.269 $. The Ising model exhibits an ordered ferromagnetic phase when $T<T_c$, and a disordered paramagnetic phase when $T>T_c$. 

For the macroscopic observables, we consider the magnetization and domain wall density defined by:
\begin{equation}
    \rho_{\text{DW}} = \frac{1}{4L^2} \sum_{\langle i,j \rangle} (1 - \sigma_i \sigma_j) .
\end{equation}
The domain wall density quantifies the degree of disorder by measuring the fraction of misaligned nearest-neighbor spin pairs. We derive two additional closure variables using an autoencoder, resulting in a latent state $\bz$ of dimension $4$.

In the first part, we fix the large-system size to $n=64^2$, while varying the computational power of the microscopic simulator $\mathcal{S}_{n_s}$ by considering $n_s \in \{ 8^2, 16^2, 32 ^2, 64^2\}$. 
When $n_s = 64^2$, our method reduces to traditional methods for deriving the macroscopic dynamics. We set the external field to $ h=0.1$ and the temperature to $T=2.5>T_c$.  To generate the large-system dataset $D$, we employ the hierarchical algorithm with $\log_2(n/n_s)$ iterations. For each iteration, we first perform an \textsc{Upsample} step to replicate every spin into $2^2$ block. Next, we perform a \textsc{LocalRelax} step by evolving a short-time continuous-time Glauber dynamics locally to remove the unphysical artifacts introduced in the \textsc{Upsample} step.

\cref{fig:Ising_scale} and \cref{fig:ising_traj} compare the performance of our method with the baseline.
From \cref{fig:Ising_scale}, we observe that our method consistently outperforms the baseline when $n_s< n$, which demonstrates the effectiveness of our method. 
\cref{fig:ising_traj} further compares the magnetization statistics and domain wall density statistics derived from ground truth trajectories with those produced by the baseline model and our method.
The statistics of the trajectories predicted by our method align closely with the ground truth, whereas the baseline exhibits much larger variations.

The parameter $n_s$ can influence our method in two ways: 
\textit{(i)} The size of the small-system dataset $D_s$ is $n_s$. For smaller $n_s$, more iterations of upsampling are required to obtain $D$, which may degrade the data quality of $D$. 
\textit{(ii)} When generating $\bx_{t+\delta t, \mathcal{I}}$ from $\bx\sim D$, we perform partial evolution on a small patch of size $n_s$. A smaller $n_s$ leads to higher stochasticity of the generated data. As a result, a smaller $n_s$ generally leads to worse performance. 
However, the test error of our method remains relatively low when $n_s \geq 16^2$, which demonstrates the robustness of our method \textit{w.r.t.} $n_s$.

\begin{figure*}[t]
    \centering
    \begin{tabular}{@{}c@{\hspace{5pt}}@{\hspace{5pt}}c@{\hspace{2pt}}c@{}}
        \quad \ \  {(a)  Microscopic Configuration}
         & \ \ \
         {(b) Order Parameters vs. Temperature}
         &  \ \ 
         {(c) Macroscopic Dynamics}\\
         \includegraphics[width=0.2\textwidth]{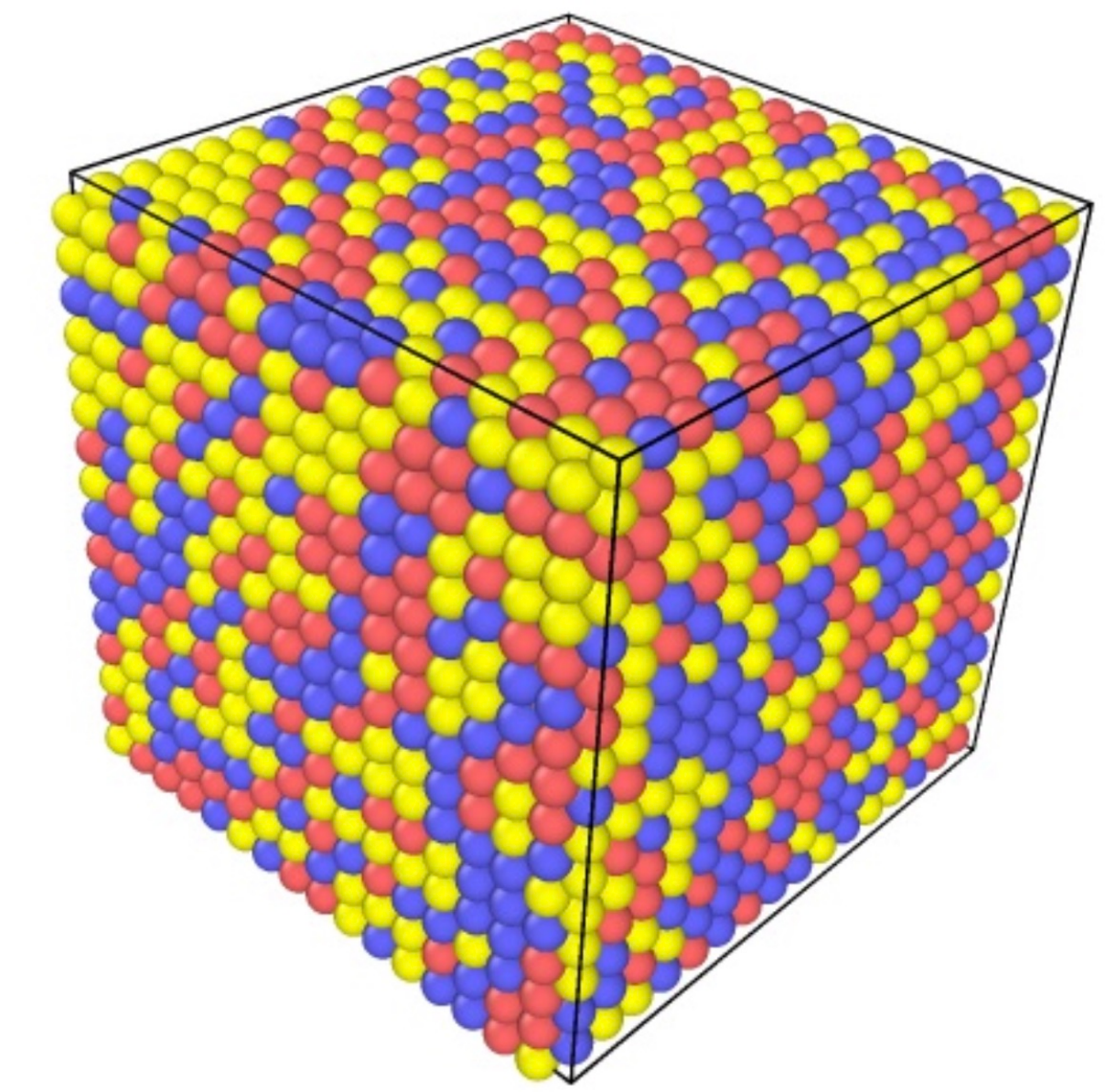} & \includegraphics[width=0.32\textwidth]{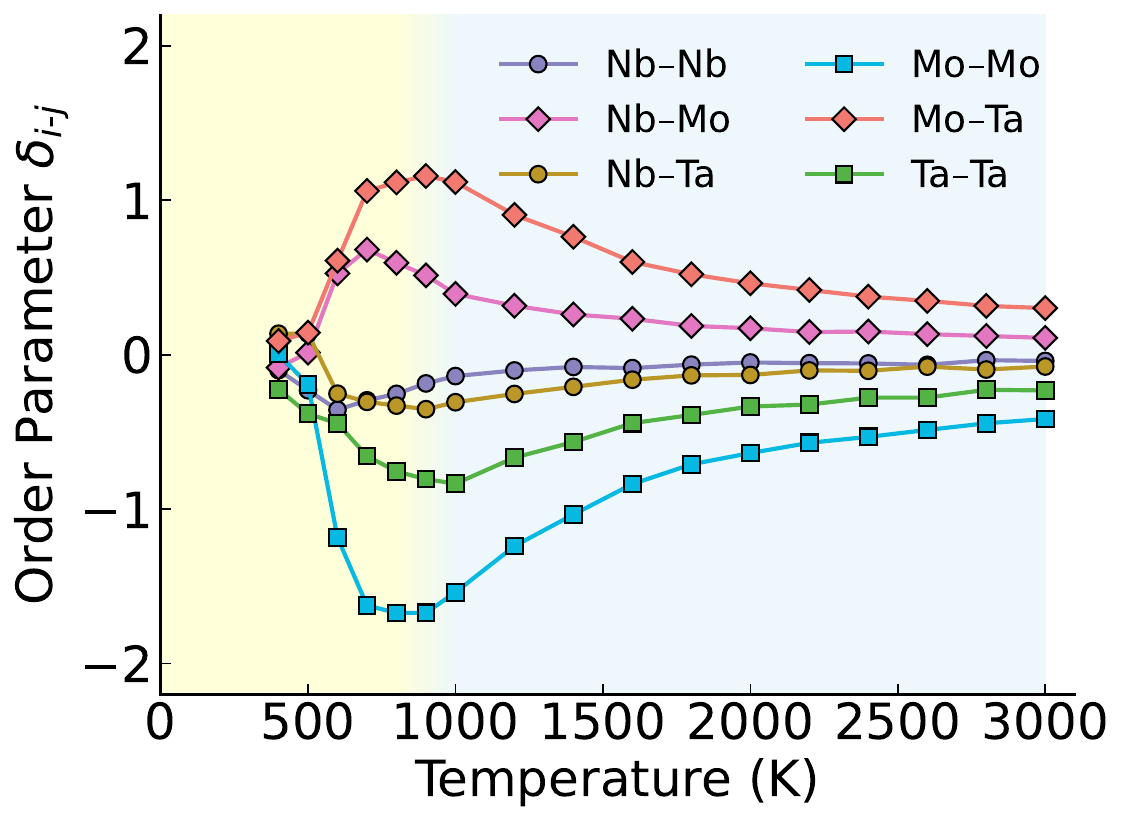} &
         \includegraphics[width=0.38\textwidth]{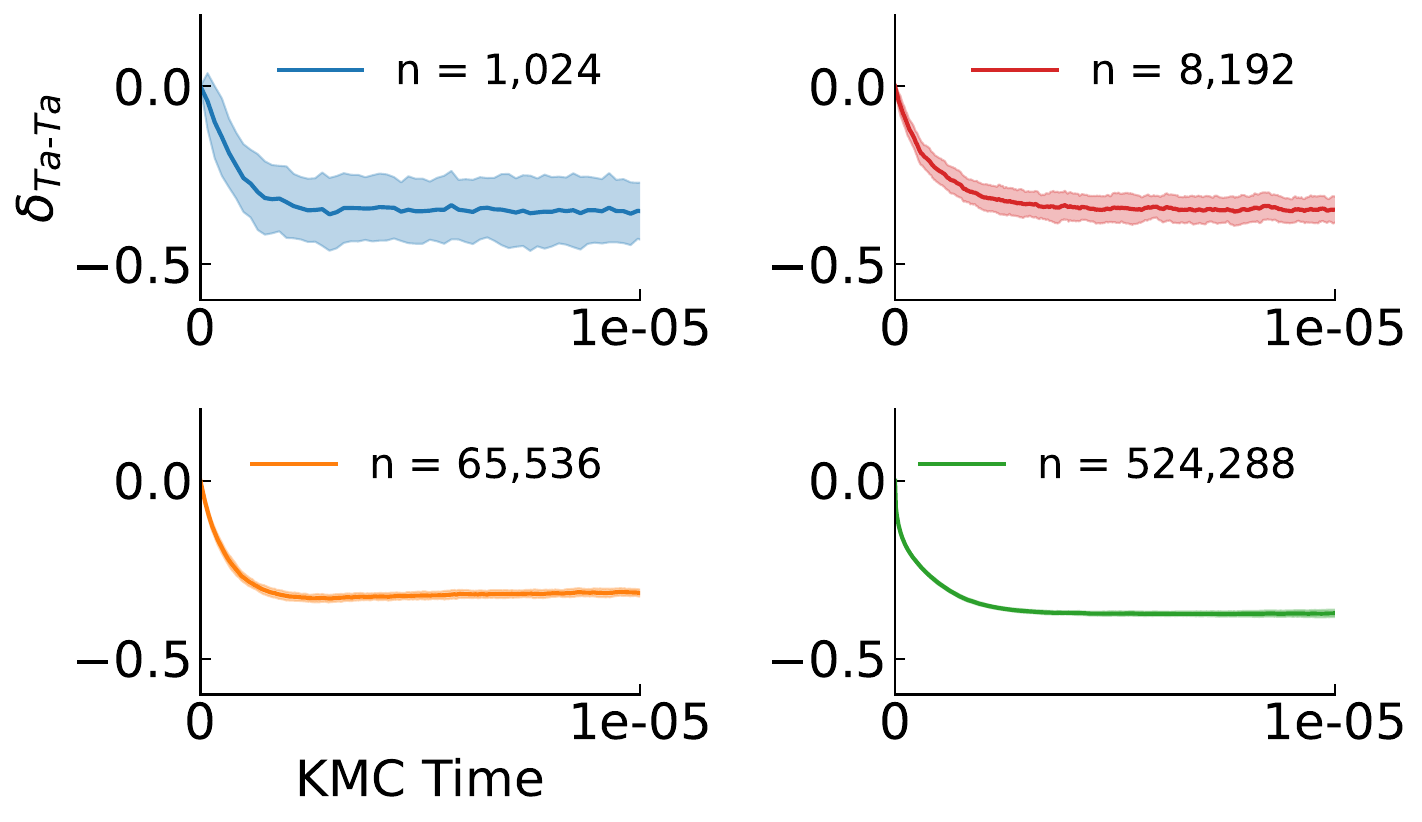}\vspace{-5pt}
         \\
    \end{tabular}
    \caption{Results of the NbMoTa equimolar alloy system. (a) Microscopic configuration of NbMoTa alloy with $8,192$ atoms. (b) Equilibrium order parameters as a function of temperature, obtained from macroscopic dynamics simulations by the trained SDE model. The microscopic system contains $n=8192$ atoms. (c) Macroscopic dynamics of $\delta_{\text{Ta-Ta}}$  for microscopic systems of varying sizes when $T=2000K$. The mean and standard deviations are calculated over 100 trajectories. 
    }
    \label{fig:alloy}
\end{figure*}

In the second part of the experiment, we demonstrate that our method is capable of capturing the critical behavior of macroscopic dynamics and estimating the critical exponents. 
Finite-size scaling theory describes how equilibrium observables of a finite system with size $n=L\times L$ scale with $L$ and $T$ near the critical temperature. Specifically, for the equilibrium magnetization $\langle |M| \rangle$ and magnetic susceptibility $\mathcal{X}$, we will have~\cite{sandvik2010computational}:
\begin{equation}\label{eq:finite_size_scaling}
    \begin{aligned}
        \langle |M| \rangle (T, L) &= L^{-\beta / \nu}  \mathcal{F}_{M}\left((T-T_c) L^{1/\nu}\right) , \\
        \mathcal{X} (T, L) &= L^{\gamma/\nu} \mathcal{F}_{\mathcal{X}} \left( (T-T_c) L^{1/\nu} \right) , 
    \end{aligned}
\end{equation}
where $\mathcal{F}_M$ and $\mathcal{F}_{\mathcal{X}}$ are scaling functions and the magnetic susceptibility $\mathcal{X}$ is defined as:
\begin{equation}
    \mathcal{X} = \frac{L^2}{T}\left( \langle M^2\rangle - \langle |M|^2 \rangle \right).
\end{equation}
When $T = T_c$, \cref{eq:finite_size_scaling} simplify to:
\begin{equation}\label{eq:critical_scaling}
    \begin{aligned}
        \langle |M| \rangle (T_c, L) & \sim L^{-\beta/\nu} , \\
        \mathcal{X} (T_c, L) &\sim L^{\gamma/\nu}.
    \end{aligned}
\end{equation}
For the two-dimensional Ising model, the theoretical values of the critical exponents are $\beta = 1/8, \nu=1, \gamma=7/4$~\cite{sandvik2010computational}. 

In this part of the experiment, we fix the small-system size to be $n_s = 16^2$, while varying the large-system size $n=L \times L$. We apply our method across different system sizes and a range of temperatures near $T_c$.
Next, we perform long-time simulations towards equilibrium using the trained SDE models.
We calculate the equilibrium magnetization and magnetic susceptibility from the predicted trajectories,
to which we fit the critical exponents.

In \cref{fig:ising_scaling} (a) and (b), we plot the equilibrium magnetization and susceptibility as a function of $L$ at $T=2.27$, which is very close to $T_c$. Next, we fit log-log curves to the data by \cref{eq:critical_scaling}, yielding  estimated critical exponent $(\beta/\nu)^{\ast} = 0.129, (\gamma/\nu)^{\ast} =1.670$. The estimated critical exponents are very close to the theoretical value $\beta/\nu = 0.125, \gamma/\nu = 1.75$. 
These results show that our method can accurately recover the ratios $\beta/\nu$ and $\gamma/\nu$, demonstrating its ability to capture accurate macroscopic dynamics across system sizes.   
This is especially significant because we only use microscopic simulations of small systems with size $n_s=16^2$, yet our method still succeeds in reproducing macroscopic dynamics for large systems with different sizes.

We further evaluate whether our method can recover the individual critical exponents $\beta, \nu, \gamma$. 
To achieve this, we estimate $\nu$ using the Binder cumulant, defined as: 
\begin{equation}
    U(T, L) = 1 - \frac{\langle M^4 \rangle}{3 \langle M^2 \rangle^2}.
\end{equation}
According to finite-size scaling theory, the slope of the Binder cumulant at the critical temperature will scale with $L$ as follows~\cite{sandvik2010computational}: 
\begin{equation}
    |dU/dT|_{T=T_c} \sim L^{1/\nu} .
\end{equation}
We plot the raw Binder cumulant data in \cref{fig:ising_scaling} (c) and compute the slope at $T_c$ by fitting a curve to the data. In \cref{fig:ising_scaling} (d), a log-log fit gives the  estimated $\nu^{\ast} = 0.97$, which is close to the theoretical value of $\nu = 1$. 

From \cref{eq:finite_size_scaling}, we note that if we plot $\langle |M| \rangle L^{\beta/\nu}$ as a function of $(T-T_c) L^{1/\nu}$, the data should collapse onto the same curve for different $T$ and $L$. Similarly, $\mathcal{X}L^{-\gamma/\nu}$ and $(T-T_c)L^{1/\nu}$ should also collapse onto the same curve. We present these results in \cref{fig:finite_size_scaling}. We can observe that even though the raw magnetization data and raw susceptibility data lie on different curves for different $T$ and $L$, the scaled data approximately collapse onto the same curve. This validates that our method can accurately capture the critical behavior of macroscopic dynamics in the Ising model.

\subsection{NbMoTa alloy}\label{exp:alloy}
We next validate our method on a more realistic NbMoTa equimolar alloy system to demonstrate its robustness in handling complex microscopic systems. We adopt the experimental setting of Ref.~\cite{xingNeuralNetworkKinetics2024}, where a neural network was trained to capture energy barriers and used to investigate microscopic diffusion dynamics.

The microscopic diffusion dynamics are modeled using the Kinetic Monte Carlo (KMC) algorithm. 
In body-centered cubic (BCC) systems such as NbMoTa alloy, each vacancy $i$ can jump to one of its eight first-nearest neighbors with rate $k_{ij} = k_0 \exp(-E_{ij}/k_BT), j = 1, \cdots 8$, where $ E_{ij} $ is the energy barrier for the jump to the $j$-th neighbor and $k_0$ is an attempt frequency. 
In our study, we employ the pretrained neural network model from Ref.~\cite{xingNeuralNetworkKinetics2024} for energy barrier calculation. 
At each KMC step, the transition rates for all vacancies are computed, giving the total jump rate $R_{\text{tot}} = \sum_{i, j} k_{ij}$.
Next, the  time increment is sampled from the exponential distribution $\Delta t \sim -\frac{\ln r}{R_{\text{tot}}}, r\sim \mathcal{U}(0, 1)$. For large systems, the number of vacancies is substantial, and evaluating transition rates for all possible events is computationally demanding.

For the macroscopic observables, we investigate the non-proportional short-range order (SRO) parameters $\delta_{\text{Nb-Nb}}, \delta_{\text{Nb-Mo}}, \delta_{\text{Nb-Ta}}, \delta_{\text{Mo-Mo}}, \delta_{\text{Mo-Ta}},\delta_{\text{Ta-Ta}}$. The SRO $\delta_{ij}$ is defined by:
\begin{equation}
    \delta_{i\text{-}j} = \frac{n_{ij} - n_{0,ij}}{n_i} ,
\end{equation}
where $n_i$ is the number of atoms of type $i$, $n_{ij}$ denotes the number of pairs between atom $i$ and $j$ in the first-nearest neighbor shell, and $n_{0, ij}$ denotes the number of pairs in random solutions. Since we consider an equimolar NbMoTa alloy with BCC structure, $n_{0, ij}=8/3$. For random configuration, $\delta_{i\text{-}j} = 0 $. A positive $\delta_{i\text{-}j}$ indicates a favored $i\text{-}j$ pair, while a negative $\delta_{i\text{-}j}$ indicates an unfavored $i\text{-}j$ pair. The order parameter $\delta_{i\text{-}j}$ is essentially a rescaled form of the well-studied Warren–Cowley SRO parameter~\cite{cowley1965short, fernandez2017short,han2024ubiquitous}. 

In our experiment, we set the small-system size to  $n_s = 1,024$, with a vacancy concentration corresponding to $1$ vacancy per $1024$ sites.  Denote the supercell length by $L$, then the supercell will consist of $2L^3$ atoms for a BCC lattice. Hence, the supercell length of the small system is $L=8$. 
Within the hierarchical upsampling scheme, an \textsc{Upsample} step is performed by concatenating multiple small-system configurations, followed by a \textsc{LocalRelax} step where KMC dynamics are run within small patches to remove unphysical artifacts. 
We first train an SDE model for the small system with $n_s=1024$ atoms, and find that learning the macroscopic dynamics of the $6$ macroscopic observables directly can already give accurate results consistent with microscopic simulations. 
Therefore, for the macroscopic dynamics derivation of the large system from data $D$, no closure variables are derived and the latent dimension is $6$. 

We first consider a large system with $n=8,192$ atoms. Training data are collected across temperatures from $T=400K$ to $T=3000K$, and an SDE model dependent on $T$ is trained:
\begin{equation}
    \mathrm{d}\bz_t = \bmu(\bz_t, T)\rd t + \bSigma^{1/2}(\bz_t, T) \rd \mathbf{B}_t .
\end{equation}
The trained SDE is subsequently simulated for a long time at different temperatures. \cref{fig:alloy} (b) shows that the absolute value of $\delta_{\text{Mo-Ta}}, \delta_{\text{Ta-Ta}}, \delta_{\text{Mo-Mo}}$ increases for $T<800K$, reaches a maximum around $T=800\text{-}900K$, and decreases thereafter. 
The results indicate a critical temperature around $800\sim900K$, corresponding to the regime of maximal diffusion-favored ordering. 
Our results correspond to the results obtained from the microscopic simulation in Ref.~\cite{xingNeuralNetworkKinetics2024}, confirming the accuracy and effectiveness of our method in detecting phase transitions.

To assess scalability, we further extend our method to much larger systems with $n=65,536 (L=32)$ and $n=524,288 (L=64)$ atoms at $T=2000K$. 
In \cref{fig:alloy} (c), we show the macroscopic dynamics of $\delta_{\text{Ta-Ta}}$ for different system sizes at $T=2000K$.
While larger systems require more KMC time steps to reach equilibrium, each KMC step corresponds to a smaller time increment. From \cref{fig:alloy}(c), we observe that the equilibrium KMC time is nearly identical across systems of different sizes. While the mean of the macroscopic dynamics converges approximately to the same value, the trajectories of smaller systems are more stochastic.

\section{DISCUSSION}

This work proposes a framework to learn macroscopic dynamics of large microscopic systems from small-system simulations.
We apply our method to SPDEs, spin systems, and an NbMoTa alloy system, scaling it to a large system with $524{,}288$ atoms.
Through these applications, we highlight the capability of our model to capture accurate macroscopic dynamics over a wide range of temperatures and system sizes.

The conventional multiscale workflow typically begins by assuming a specific continuum evolution equation \emph{a priori} and then determines the associated coefficients from experiments or atomistic simulations. Common choices for the continuum evolution equation include Fick’s laws~\cite{crank1979mathematics}, Onsager transport equations~\cite{de2013non}, and phase-field models such as the Allen-Cahn and Cahn-Hilliard equations~\cite{cahn1958free, allen1979microscopic, chen2002phase}. For example, diffusion is typically described by Fick's law $\partial_t c=\nabla\!\cdot\!\big(D\nabla c\big)$, where the diffusion coefficient is subsequently estimated from the mean squared displacement of molecular dynamics trajectories~\cite{tsige2004molecular, liuFickDiffusionCoefficients2012}.

Both our workflow and the conventional multiscale workflow aim to construct effective dynamical descriptions from microscopic simulations to reduce the computational cost of tracking massive atomic degrees of freedom. 
However, our framework differs in several key respects. 
First, conventional multiscale approaches are typically formulated at a mesoscopic level, while our method learns the macroscopic dynamics. In conventional multiscale approaches, the spatially resolved field variables depend locally on microscopic coordinates. 
In contrast, our method explicitly targets the macroscopic dynamics. In our framework, the observables of interest depend globally on all the microscopic coordinates, resulting in macroscopic dynamics that evolve solely as a function of time rather than space.
While we utilize an encoder $\boldsymbol{\varphi}$ to extract latent features $\bphi(\mathbf{x}_{t, \mathcal{I}})$ from each patch, these features are aggregated to define the state $\mathbf{z}_t = \bphi(\mathbf{x}_t)$ for the whole large system. 
Our method explicitly models the dynamics of $\mathbf{z}_t$ from short-time, small-scale microscopic information, and is therefore best viewed as a macroscale modeling approach. 
Second, regarding the model structure, conventional multiscale approaches explicitly assume a predefined functional continuum evolution equation, which often limits their accuracy and ability to capture complex or non-linear phenomena. 
In contrast, our framework learns both the closure variables and the macroscopic dynamics directly from data, without prescribing a specific functional form. 
Although we employ structured architectures such as OnsagerNet, these structures serve only to enforce fundamental physical constraints, such as thermodynamic consistency, rather than to constrain the admissible functional forms of the dynamics.
By parameterizing the encoder and macroscopic dynamics with neural networks, we ensure high model expressivity. 
Consequently, our framework offers significantly greater flexibility in discovering the underlying governing laws directly from data. 

In our framework, we enforce physical consistency of the macroscopic dynamics by modeling the SDE using a stochastic OnsagerNet. The closure variables, on the other hand, are obtained through data-driven learning and may contain redundancy or lack direct physical interpretability. To reduce redundancy, one practical approach is to vary the dimension of the closure variables and identify when further increases no longer improve macroscopic prediction accuracy. In addition, physical interpretability may be improved by incorporating basic physical or symmetry constraints into the encoder, such as translation invariance for lattice systems with periodic boundary conditions, which we leave for future investigation. 

A practical limitation of the present framework arises from the hierarchical upsampling procedure used to generate large-system snapshots from small-system simulations.
The \textsc{LocalRelax} step is used to remove the unphysical artifacts introduced by the \textsc{Upsample} step. In this paper, \textsc{LocalRelax} is implemented by dividing each upsampled large-system snapshot into overlapping patches and applying short-time local dynamics evolution within each patch.
This simple procedure is effective in our experiments for removing common unphysical artifacts introduced by upsampling.
However, for more complex systems, this simple \textsc{LocalRelax} step may be insufficient to adequately remove the unphysical artifacts. 
Moreover, when applied to extremely large systems with billions of atoms, the hierarchical upsampling scheme requires many iterations, and the quality of the generated dataset $D$ may deteriorate as the number of iterations increases.
The hierarchical upsampling strategy adopted in this work represents only one possible approach, and more sophisticated strategies may be required for more complex systems.
We acknowledge this as a limitation of the current method.
In the future, we will explore more efficient strategies for generating large-system datasets $D$ to mitigate this challenge. 

Looking ahead, this framework holds significant promise for bridging the gap between microscopic simulations and macroscopic material behavior in complex material systems, such as high-entropy alloys~\cite{choi2018understanding,kostiuchenko2019impact,cao2025capturing} and polymer solutions~\cite{kremer1988monte,gartner2019modeling}.
We envision this framework as a powerful tool for accelerating the discovery and design of advanced functional materials, with potential applications in energy storage, catalysis, and structural technologies.

\begin{acknowledgments}
This research is supported by the National Research Foundation, Singapore under its AI Singapore Programme (AISG Award No: AISG3-RP-2022-028),  the Ministry of Education, Singapore, under its funding for the Research Centre of Excellence Institute for Functional Intelligent Materials (Project No. EDUNC-33-18-279-V12), and Academic Research Fund Tier 3 Grant (Project No. MOET32024-0002). 
The computational work for this article was partially performed on resources of the National Supercomputing Centre, Singapore (https://www.nscc.sg).
\end{acknowledgments}

\section*{Data Availability}
The supporting data and code are available in the public repository~\cite{chen2026code}.

\appendix
\section{Notation and Symbols}\label{app:notation}
The notation and symbols are summarized in \cref{tab:symbols}.
\begin{table*}[htbp]
\centering
\label{tab:symbols}
\renewcommand{\arraystretch}{1.12}
\begin{tabularx}{\linewidth}{@{} l @{\hspace{0.3cm}} l @{}}
\toprule
Symbol & Meaning \\
\midrule
$\mathbf{x}_t$ & Microscopic state of the large system. \\

$\mathcal{I}$ & Index set of a randomly selected patch within the large system. \\

$\delta t$ & Time step. \\

$x_{t,\mathcal{I}}$ & Microscopic state restricted to patch $\mathcal{I}$ at time $t$. \\

$x_{t+\delta t,\mathcal{I}}$ & Microscopic state restricted to patch $\mathcal{I}$ at time $t+\delta t$. \\

$n_s$ & Number of lattice sites in the small system. \\

$n$ & Number of lattice sites in the large system. \\

$\mathcal{S}_{n_s}$ & Microscopic simulator that can accurately simulate the microscopic dynamics of a small system up to $n_s$ lattice sites. \\

$D_s$ & Dataset of small-system snapshots. \\

$D$ & Training dataset constructed from $D_s$ through hierarchical upsampling scheme. \\

$K$ & Number of patches used to partition the large system. \\

$\boldsymbol{\varphi}^{\ast}$ & Mapping from a microscopic state to macroscopic observables. \\ 

$\hat{\boldsymbol{\varphi}}$ & Mapping from a microscopic state to closure variables. \\ 

$\boldsymbol{\varphi}$ & Combined encoding map, $\boldsymbol{\varphi}=(\boldsymbol{\varphi}^{\ast},\hat{\boldsymbol{\varphi}})$. \\

$\boldsymbol{\psi}$ & Decoder.  \\

$\bmu$ & Drift term of the latent dynamics. \\

$\bSigma$ & Diffusion term of the latent dynamics. \\

$\bz^{\ast}$ & Macroscopic observables defined by $\bz^{\ast} = \bphi^{\ast}(\bx)$ \\

$\hat{\bz}$ & Closure variables defined by $\hat{\bz} = \hat{\bphi}(\bx)$ \\ 

$\bz_t$ & Latent state defined by $\bz_t=\bphi(\bx_t)$. \\

$\bz_{t+\delta t,\mathcal{I}}$ & Latent state corresponding to $x_{t+\delta t,\mathcal{I}}$. \\

$\Delta \bz_{t}$ & Increment of the latent state over $\delta t$, defined by $\bz_{t+\delta t} - \bz_t$ \\

$\Delta \bz^{\ast}_{t}$ & Increment of the macroscopic observable, defined by $\bz^{\ast}_{t+\delta t} - \bz^{\ast}_t$ \\

$\Delta \bz_{t, \mathcal{I}}$ & Increment of the latent state restricted to patch $\mathcal{I}$, defined by $\bz_{t+\delta t, \mathcal{I}} - \bz_{t, \mathcal{I}}$ \\

$\Delta \bz^{\ast}_{t, \mathcal{I}}$ & Increment of the macroscopic observable restricted to patch $\mathcal{I}$, defined by $\bz^{\ast}_{t+\delta t, \mathcal{I}} - \bz^{\ast}_{t, \mathcal{I}}$ \\
\bottomrule
\end{tabularx}
\caption{Summary of key symbols and definitions.}
\end{table*}

\section{Continous-time Glauber dynamics}
The algorithm of continuous-time Glauber dynamics is summarized in \cref{alg:glauber}. 

\label{app:b}
\begin{figure}[h]
    \begin{algorithm}[H]
        \caption{Continuous-time Glauber Dynamics}
        \label{alg:glauber}
        \begin{algorithmic}[1]
            \Require 
                \Statex $\{\sigma_i\}_{i=1}^n$: Initial spins
                \Statex $\beta$: inverse temperature 
                \Statex \texttt{max\_steps}: maximum number of steps
            \State Initialize \texttt{step} $\gets 0$, \texttt{kmc\_time} $\gets 0$ 
            \While{\texttt{step} $<$ \texttt{max\_steps}}
                \For{$i = 1$ to $n$}
                    \State Compute energy differences $\Delta H_i $ by flipping spin $i$
                    \State Compute the rate $ r_i = \tfrac{1}{1 + e^{\beta \Delta H_i}}$ 
                \EndFor
                \State Compute the total rate $R = \sum_{i=1}^N r_i$
                \State Sample a spin $j$ to flip with probability $r_i / R$
                \State Sample time step $\delta t  = - \tfrac{\ln u}{R}, u\sim \mathcal{U}(0, 1)$
                \State \texttt{kmc\_time} $\gets$ \texttt{kmc\_time} + $\delta t $, 
                \State \texttt{step} $\gets$ \texttt{step} $+ 1$
            \EndWhile
        \end{algorithmic}
    \end{algorithm}
\end{figure}

\section{Theoretical Analysis}\label{app:a}

\begin{proof}[Proof of \cref{thm:1}]
We divide the proof into two parts: first, we establish the closed-form expressions for the unique minimizers using variational calculus; second, we derive the error bounds based on the total variation distance.
\paragraph*{1. Derivation of Minimizers}
The loss function $\mathcal{L}$ and $\mathcal{L}_p$ can be rewritten as:
\begin{equation*}
    \begin{aligned}
        \mathcal{L}[\bmu, \bSigma] &= \mathbb{E}_{\bx_t, \delta t} \, \mathbb{E}_{\bx_{t+\delta t }|\bx_t}
    \big[ \\
    & -2\log p \big( \bz_{t+\delta t } \,\big|\, \bz_t + \bmu(\bz_t) \delta t , \, \bSigma(\bz_t) \delta t  \big) \big] , \\
    \mathcal{L}_p[\bmu, \bSigma] &= \mathbb{E}_{\bx_t, \delta t} \, \mathbb{E}_\mathcal{I} \, \mathbb{E}_{q(\bx_{t+\delta t , \mathcal{I}}|\bx_t)}
    \big[ \\
    & -2\log p \big( \bz_{t+\delta t , \mathcal{I}} \,\big|\, \bz_t + \bmu(\bz_t) \delta t , \, K \, \bSigma(\bz_t) \delta t  \big) \big] .
    \end{aligned}
\end{equation*}
Recall $\Delta \bz_t= \bz_{t+\delta t} - \bz_t, \Delta \bz_{t, \mathcal{I}} = \bz_{t+\delta t, \mathcal{I}} - \bz_t$. 
By introducing the precision matrix $\bLambda = \bSigma^{-1}$ and the displacement vector 
\begin{align*}
    &\bdelta(\bz_t, \bz_{t+\delta t }) \coloneqq \Delta \bz_t - \bmu(\bz_t) \delta t  \\
    &\bdelta_\mathcal{I}(\bz_t, \bz_{t+\delta t , \mathcal{I}}) \coloneqq \Delta \bz_{t, \mathcal{I}} - \bmu(\bz_t) \delta t, 
\end{align*} 
we can rewrite the objective as:  
\begin{equation*}
    \begin{aligned}
    \mathcal{L}[\bmu, \bSigma]
    &=\mathbb{E}_{\bx_t, \delta t} \mathbb{E}_{\bx_{t+\delta t }|\bx_t} \big[  d \log (2\pi) + \log |\bSigma(\bz_t)\delta t | \big] \\
    & + \bdelta(\bz_t, \bz_{t+\delta t })^T (\bSigma(\bz_t) \delta t ) ^{-1} \bdelta(\bz_t, \bz_{t+\delta t }), \\
     \mathcal{L}_p[\bmu, \bSigma]&= \mathbb{E}_{\bx_t, \delta t}  \mathbb{E}_\mathcal{I} \mathbb{E}_{q(\bx_{t+\delta t , \mathcal{I}}|\bx_t)} \big[  
    d \log (2\pi) \\
 &+ \log |K\bSigma(\bz_t)\delta t | \\
 & + \bdelta_\mathcal{I}(\bz_t, \bz_{t+\delta t , \mathcal{I}})^T (K\bSigma(\bz_t) \delta t ) ^{-1} \bdelta_\mathcal{I}(\bz_t, \bz_{t+\delta t , \mathcal{I}}) \big].
    \end{aligned}
\end{equation*}
For further simplicity, we omit the notational dependency of $\bdelta, \bdelta_{\mathcal{I}}, \bmu, \Lambda$ on $\bz_t, \bz_{t+\delta t }, \bz_{t+\delta t,\mathcal{I}}$. We will proceed to calculate the first-order variation and second-order variation of $\hat{\mathcal{L}}$: 
\begin{align*}
&\quad \hat{\mathcal{L}}[\bmu + \epsilon \bh, \bLambda + \epsilon \mathbf{H}] \\ 
&=\mathbb{E}_{\bx_t, \delta t} \mathbb{E}_{\bx_{t+\delta t }|\bx_t} \big[  d \log (2\pi) + \log(\delta t )- \log |\bLambda + \epsilon \mathbf{H}|  \\
&\quad + \tfrac{1}{\delta t }(\bdelta - \epsilon \bh \delta t  )^T (\bLambda + \epsilon \mathbf{H})(\bdelta  - \epsilon \bh\delta t  ) \big] \\
& = \hat{\mathcal{L}}[\bmu, \bLambda] -   \mathbb{E}_{\bx_t, \delta t} \mathbb{E}_{\bx_{t+\delta t }|\bx_t}\big[ \log |\bLambda + \epsilon \mathbf{H}|  - \log |\bLambda|  \big] \\
&\quad + \epsilon \mathbb{E}_{\bx_t, \delta t}\mathbb{E}_{\bx_{t+\delta t }|\bx_t}\big[\tfrac{1}{\delta t }\bdelta^T\mathbf{H}\bdelta - 2\bdelta^T\bLambda\bh \big] \\
&\quad + \tfrac{\epsilon^2}{2} \mathbb{E}_{\bx_t, \delta t}\mathbb{E}_{\bx_{t+\delta t }|\bx_t} \big[2\delta t  \bh^T\bLambda \bh - 4\bdelta^T \mathbf{H} \bh \big] +  o(\epsilon^2). 
\end{align*}
Since 
\begin{align*}
    \log |\bLambda + \epsilon \mathbf{H}|  &= \log |\bLambda| + \epsilon  \text{tr}(\bLambda^{-1} \mathbf{H}) \\
    &- \tfrac{\epsilon^2}{2} \text{tr}(\bLambda^{-1} \mathbf{H}\bLambda^{-1} \mathbf{H}) 
    + o(\epsilon^2), 
\end{align*}
\begin{align*}
&\quad \hat{\mathcal{L}}[\bmu + \epsilon \bh, \bLambda + \epsilon \mathbf{H}]  \\
&= \hat{\mathcal{L}}[\bmu, \bLambda] \\
&+ \epsilon  \mathbb{E}_{\bx_t, \delta t}\mathbb{E}_{\bx_{t+\delta t }|\bx_t} \big[-\text{tr}(\bLambda^{-1} \mathbf{H}) + \tfrac{1}{\delta t }\bdelta^T\mathbf{H}\bdelta - 2\bdelta^T\bLambda\bh\big] \\
&+ \tfrac{\epsilon^2}{2} \mathbb{E}_{\bx_t, \delta t}\mathbb{E}_{\bx_{t+\delta t }|\bx_t}  \big[\text{tr}(\bLambda^{-1} \mathbf{H}\bLambda^{-1} \mathbf{H}) \\
&\quad + 2\delta t  \bh^T\bLambda \bh - 4\bdelta^T \mathbf{H} \bh\big] \\
&+  o(\epsilon^2). 
\end{align*}
Hence we have the first variation $\delta \hat{\mathcal{L}}$ and second order variation $\delta^2 \hat{\mathcal{L}}$:
\begin{equation*}
\begin{aligned}
    \delta \hat{\mathcal{L}} &= \mathbb{E}_{\bx_t, \delta t}\mathbb{E}_{\bx_{t+\delta t }|\bx_t}  \big [-\text{tr}(\bLambda^{-1} \mathbf{H}) \\
    &+ \tfrac{1}{\delta t }\bdelta^T\mathbf{H}\bdelta - 2\bdelta^T\bLambda\bh \big],  \\
    \delta^2 \hat{\mathcal{L}} &=\mathbb{E}_{\bx_t, \delta t}\mathbb{E}_{\bx_{t+\delta t }|\bx_t}  \big[\text{tr}(\bLambda^{-1} \mathbf{H}\bLambda^{-1} \mathbf{H})
    \\
    &+ 2\delta t  \bh^T\bLambda \bh - 4\bdelta^T \mathbf{H} \bh \big].
\end{aligned}
\end{equation*} 
If $\bmu^{\ast}, \bLambda^{\ast}$ are the minimizer of $\hat{\mathcal{L}}$, we must have $\delta \hat{\mathcal{L}} = 0$ for all the admissible $\bh$ and $\mathbf{H}$. By a direct calculation:
\begin{align*}
     \delta \hat{\mathcal{L}} &= \mathbb{E}_{\bx_t, \delta t}\mathbb{E}_{\bx_{t+\delta t }|\bx_t}  \big[-\text{tr}(\bLambda^{-1} \mathbf{H}) + \tfrac{1}{\delta t }\bdelta^T\mathbf{H}\bdelta - 2\bdelta^T\bLambda\bh \big] \\ 
&= \mathbb{E}_{\bx_t, \delta t} \Big[  \text{tr}\big( \mathbb{E}_{\bx_{t+\delta t }|\bx_t} \big[  \tfrac{1}{\delta t } \bdelta \bdelta^T - \bLambda^{-1}\big] \mathbf{H} \big) \Big] \\
&\quad + \mathbb{E}_{\bx_t, \delta t} \Big[ \mathbb{E}_{\bx_{t+\delta t }|\bx_t} \big[ - 2\bdelta^T\bLambda\big] \bh \Big], 
\end{align*}
The stationary condition implies: 
\begin{align*}
    &\quad \mathbb{E}_{\bx_{t+\delta t }|\bx_t} \big[  \tfrac{1}{\delta t } \bdelta^{\ast} (\bdelta^{\ast})^T - (\bLambda^{\ast})^{-1}\big] =0 \quad \textit{a.e.} , \\
    &\quad \mathbb{E}_{\bx_{t+\delta t }|\bx_t} \big[ - 2(\bdelta^{\ast})^T\bLambda \big]  \\
    &= \mathbb{E}_{\bx_{t+\delta t }|\bx_t} \big[ - 2 (\Delta \bz_t- \bmu^{\ast}(\bz_t) \delta t  )^T \big] \bLambda^{\ast}(\bz_t)  \\
    &= 0 \quad  \textit{a.e.} , 
\end{align*}
where $\boldsymbol{\delta^{\ast}}(\bz_t, \bz_{t+\delta t}) \coloneqq \Delta \bz_t -\bmu^{\ast}(\bz_t) \delta t  $.  Solving these yields the unique minimizers for $\mathcal{L}$:
\begin{align*}
    \bmu^{\ast}(\bz_t) &=\tfrac{1}{\delta t } \mathbb{E}_{\bx_{t+\delta t }|\bx_t} [\Delta \bz_t] = \tfrac{1}{\delta t } \mathbb{E}_{\bx_{t+\delta t }|\bx_t} [\bz_{t+\delta t } - \bz_t] \quad \textit{a.e.} , \\ 
   \bSigma^{\ast}(\bz_t)&=( \bLambda^{\ast})^{-1}(\bz_t) =   \tfrac{1}{\delta t }\mathbb{E}_{\bx_{t+\delta t }|\bx_t} [ \bdelta^{\ast} (\bdelta^{\ast})^T ]
   \quad \textit{a.e.} . 
\end{align*}
We can calculate the second variation at $\bmu^{\ast}, \bLambda^{\ast}$:
\begin{align*}
     \delta^2 \hat{\mathcal{L}} &= \mathbb{E}_{\bx_t, \delta t}\mathbb{E}_{\bx_{t+\delta t }|\bx_t}  \big[\text{tr}(\bLambda^{-1} \bH\bLambda^{-1} \bH) + 2\delta t  \bh^T\bLambda \bh\big] \\
     &\quad -4 \mathbb{E}_{\bx_t, \delta t} \Big[ \mathbb{E}_{\bx_{t+\delta t }|\bx_t} \big[(\Delta \bz_t - \bmu^{\ast}(\bz_t)\delta t )^T\big] \bH \bh \Big] \\ 
     &= \mathbb{E}_{\bx_t, \delta t}\mathbb{E}_{\bx_{t+\delta t }|\bx_t}  \big[\text{tr}(\bLambda^{-1} \bH\bLambda^{-1} \bH) + 2\delta t  \bh^T\bLambda \bh\big] ,
\end{align*}
$ \delta^2 \hat{\mathcal{L}}>0$ if either $\bH$ or $\bh$ is not equal to zero almost everywhere. 
Then $\bmu^{\ast}$ and $\bLambda^{\ast}$ are the unique minimizer of functional $\hat{\mathcal{L}}$. Equivalently, $\bmu^{\ast}$ and $\bSigma^{\ast}$ are the unique minimizer of $\mathcal{L}$. 

Similarly, we can compute the first-order variation and second-order variation of $\hat{\mathcal{L}}_p$: 
\begin{align*}
    \hat{\mathcal{L}}_p[\bmu + \epsilon \bh, \bLambda + \epsilon\bH] = \hat{\mathcal{L}}_p[\bmu, \bLambda] + \epsilon \delta \hat{\mathcal{L}}_p + \tfrac{\epsilon^2}{2} \delta^2 \hat{\mathcal{L}}_p + o(\epsilon^2) ,
\end{align*}
where 
\begin{align*}
    \delta \hat{\mathcal{L}}_p &= \mathbb{E}_{\bx_t, \delta t} \mathbb{E}_\mathcal{I}\mathbb{E}_{q(\bx_{t+\delta t , \mathcal{I}}|\bx_t)}  \big [-\text{tr}(\bLambda^{-1} \bH) \\
    &\quad + \tfrac{1}{K\delta t}\bdelta_\mathcal{I}^T \bH \bdelta_\mathcal{I} - \tfrac{2}{K}\bdelta_\mathcal{I}^T\bLambda\bh \big], 
\end{align*}
\begin{align*}
    \delta^2 \hat{\mathcal{L}}_p &=  \mathbb{E}_{\bx_t, \delta t} \mathbb{E}_\mathcal{I}\mathbb{E}_{q(\bx_{t+\delta t , \mathcal{I}}|\bx_t)}  \big[\text{tr}(\bLambda^{-1} \bH\bLambda^{-1} \bH)\\
    &\quad + \tfrac{2\delta t }{K} \bh^T\bLambda \bh - \tfrac{4}{K}\bdelta_\mathcal{I}^T \bH \bh \big]. 
\end{align*} 

If $\bmu^{\dagger}, \bLambda^{\dagger}$ are the minimizer of $\hat{\mathcal{L}}_p$, we must have $\delta \hat{\mathcal{L}}_p = 0$ for all the admissible $\bh$ and $\bH$. Let $\bdelta_{\mathcal{I}}^{\dagger}(\bz_t, \bz_{t+\delta, \mathcal{I}})\coloneqq \Delta \bz_{t, \mathcal{I}}- \bmu^{\dagger}(\bz_t)\delta t$, since
\begin{align*}
     \delta \hat{\mathcal{L}}_p &=  \mathbb{E}_{\bx_t, \delta t} \mathbb{E}_\mathcal{I}\mathbb{E}_{q(\bx_{t+\delta t , \mathcal{I}}|\bx_t)}   \big [-\text{tr}(\bLambda^{-1} \bH) \\
     &\quad+ \tfrac{1}{K\delta t}\bdelta_\mathcal{I}^T \bH \bdelta_\mathcal{I} - \tfrac{2}{K}\bdelta_\mathcal{I}^T\bLambda \bh \big] \\ 
    &= \mathbb{E}_{\bx_t, \delta t} \Big[  \text{tr}\big( \mathbb{E}_\mathcal{I} \mathbb{E}_{q(\bx_{t+\delta t , \mathcal{I}}|\bx_t)} \big[  \tfrac{1}{K\delta t} \bdelta_\mathcal{I} \bdelta_\mathcal{I}^T - \bLambda^{-1}\big] \bH \big) \Big] \\
    &\quad + \mathbb{E}_{\bx_t, \delta t} \Big[  \mathbb{E}_\mathcal{I} \mathbb{E}_{q(\bx_{t+\delta t , \mathcal{I}}|\bx_t)} \big[ - \tfrac{2}{K}\bdelta_\mathcal{I}^T\bLambda\big] \bh \Big], 
\end{align*}
we have 
\begin{align*}
     &\quad \mathbb{E}_\mathcal{I} \mathbb{E}_{q(\bx_{t+\delta t , \mathcal{I}}|\bx_t)} [ - \tfrac{2}{K}(\bdelta_\mathcal{I}^{\dagger})^T\bLambda^{\dagger}] \\
     &= \mathbb{E}_\mathcal{I} \mathbb{E}_{q(\bx_{t+\delta t , \mathcal{I}}|\bx_t)}  [ - \tfrac{2}{K} (\Delta \bz_{t, \mathcal{I}}- \bmu^{\dagger}(\bz_t) \delta t  )^T] \bLambda^{\dagger}(\bz_t) \\
     &=0 \quad \textit{a.e.} \\
     &\quad \mathbb{E}_\mathcal{I} \mathbb{E}_{q(\bx_{t+\delta t , \mathcal{I}}|\bx_t)}  [  \tfrac{1}{K\delta t} \bdelta_\mathcal{I}^{\dagger}(\bdelta_\mathcal{I}^{\dagger})^T - (\bLambda^{\dagger})^{-1}] =0 \quad \textit{a.e.} .
\end{align*}
hence 
\begin{align*}
    \bmu^{\dagger}(\bz_t) &= \tfrac{1}{\delta t  }\mathbb{E}_\mathcal{I} \mathbb{E}_{q(\bx_{t+\delta t , \mathcal{I}}|\bx_t)}  [ \Delta \bz_{t, \mathcal{I}} ]   \\
    &=\tfrac{1}{\delta t  }\mathbb{E}_\mathcal{I} \mathbb{E}_{q(\bx_{t+\delta t , \mathcal{I}}|\bx_t)}  [ \bz_{t+\delta t , \mathcal{I}} - \bz_{t} ] \quad \textit{a.e.}  , \\
    \bSigma^{\dagger}(\bz_t)&=( \bLambda^{\dagger})^{-1}(\bz_t) = \tfrac{1}{K\delta t} \mathbb{E}_\mathcal{I} \mathbb{E}_{q(\bx_{t+\delta t , \mathcal{I}}|\bx_t)}  [  \bdelta_\mathcal{I}^{\dagger}(\bdelta_\mathcal{I}^{\dagger})^T ] \quad \textit{a.e.} .
\end{align*}
We can calculate the second variation at $\bmu^{\dagger}, \bLambda^{\dagger}$:
\begin{align*}
    \delta^2 \hat{\mathcal{L}}_p &=  \mathbb{E}_{\bx_t, \delta t} \mathbb{E}_\mathcal{I}\mathbb{E}_{q(\bx_{t+\delta t , \mathcal{I}}|\bx_t)}  \big[\text{tr}(\bLambda^{-1} \bH\bLambda^{-1} \bH) + \tfrac{2\delta t }{K} \bh^T\bLambda^{\dagger} \bh\big] \\
    &\quad -\tfrac{4}{K}  \mathbb{E}_{\bx_t, \delta t} \mathbb{E}_\mathcal{I}\mathbb{E}_{q(\bx_{t+\delta t , \mathcal{I}}|\bx_t)}  [\bdelta^T_\mathcal{I}\bH\bh] \\
    &=\mathbb{E}_{\bx_t, \delta t} \mathbb{E}_\mathcal{I}\mathbb{E}_{q(\bx_{t+\delta t , \mathcal{I}}|\bx_t)}  \big[\text{tr}(\bLambda^{-1} \bH\bLambda^{-1} \bH) + \tfrac{2\delta t }{K} \bh^T\bLambda^{\dagger} \bh\big], 
\end{align*}

\paragraph*{2. Error Analysis}
We now bound the difference between the estimators. 
To bound $\|\bmu^{\ast} - \bmu^{\dagger}\|_{\infty}$, we use the property of Total Variation distance. For any random variable $X$ bounded by $B$, we have $$\|\mathbb{E}_q[X] - \mathbb{E}_{\hat{q}}[X]\| \le 2B \delta_{\mathrm{TV}}(q, \hat{q}).$$
Next, since the encoder $\boldsymbol{\varphi}$ is uniformly bounded by $M$, it follows that
$\|\Delta \bz_{t, \mathcal{I}}\|_{\infty} \le 2M$.
We therefore obtain: 
\begin{align*}
    &\quad\lVert \bmu^{\dagger}(\bz_t)  - \bmu^{\ast}(\bz_t)  \rVert \\
    & = \lVert \tfrac{1}{\delta t }\textstyle \mathbb{E}_{\mathcal{I}}\mathbb{E}_{q} [ \Delta \bz_{t,\mathcal{I}}]- \tfrac{1}{\delta t }\textstyle \mathbb{E}_{\bx_{t+\delta t}\mid \bx_t} [ \Delta \bz_{t}]  \rVert  \\
    &\leq \lVert \tfrac{1}{\delta t }\textstyle \mathbb{E}_{\mathcal{I}}\mathbb{E}_{q} [ \Delta \bz_{t,\mathcal{I}}]- \tfrac{1}{\delta t }\textstyle \mathbb{E}_{\bx_{t+\delta t}\mid \bx_t} [\tfrac{1}{K}\sum_{\mathcal{I}} \Delta \bz_{t,\mathcal{I}}]  \rVert  \\
    &\textstyle+ \frac{1}{\delta t} \lVert \mathbb{E}_{\bx_{t+\delta t}\mid \bx_{t}} [\Delta \bz_t - \frac{1}{K}\sum_{\mathcal{I}}\Delta \bz_{t, \mathcal{I}}] \rVert \\
    &=\lVert \tfrac{1}{\delta t }\textstyle \mathbb{E}_{\mathcal{I}}\mathbb{E}_{q} [ \Delta \bz_{t,\mathcal{I}}]- \tfrac{1}{\delta t }\textstyle \mathbb{E}_{\mathcal{I}}\mathbb{E}_{\hat{q}} [ \Delta \bz_{t,\mathcal{I}}]  \rVert \\
    &\textstyle+ \frac{1}{\delta t} \lVert \mathbb{E}_{\bx_{t+\delta t}\mid \bx_{t}} [\Delta \bz_t - \frac{1}{K}\sum_{\mathcal{I}}\Delta \bz_{t, \mathcal{I}}] \rVert 
\end{align*}
By the definition of the closure variables, we have
$
\Delta \hat{\bz}_{t}
= \textstyle \frac{1}{K}\sum_{\mathcal{I}} \Delta \hat{\bz}_{t,\mathcal{I}} .
$
By the assumption, 
\begin{equation*}
\textstyle
\bigl\| \mathbb{E}_{\bx_{t+\delta t}\mid \bx_t} [
\Delta \bz_t^{\ast}
- \tfrac{1}{K}\sum_{\mathcal{I}} \Delta \bz^{\ast}_{t,\mathcal{I}}]
\bigr\|
\le C_2 \delta t^2 ,
\end{equation*}
then 
\begin{align*}
&\quad \textstyle \frac{1}{\delta t} \lVert \mathbb{E}_{\bx_{t+\delta t}\mid \bx_{t}} [\Delta \bz_t - \frac{1}{K}\sum_{\mathcal{I}}\Delta \bz_{t, \mathcal{I}}] \rVert \\
&= \textstyle \frac{1}{\delta t} \lVert \mathbb{E}_{\bx_{t+\delta t}\mid \bx_{t}} [\Delta \bz^{\ast}_t - \frac{1}{K}\sum_{\mathcal{I}}\Delta \bz^{\ast}_{t, \mathcal{I}}] \rVert \\
&\leq C_2 \delta t 
\end{align*}
By the property of total variation distance, 
\begin{align*}
    &\quad \lVert \tfrac{1}{\delta t }\textstyle \mathbb{E}_{\mathcal{I}}\mathbb{E}_{q} [ \Delta \bz_{t,\mathcal{I}}]- \tfrac{1}{\delta t }\textstyle \mathbb{E}_{\mathcal{I}}\mathbb{E}_{\hat{q}} [ \Delta \bz_{t,\mathcal{I}}]  \rVert \\
    &\textstyle \leq \frac{1}{\delta t} \cdot 4M \cdot  \delta_{\text{TV}}(q, \hat{q}) \\
    &\leq 4MC_1 \delta t 
\end{align*}
Hence,
\[
\|\bmu^{\ast} - \bmu^{\dagger}\|_{\infty}
\le (4MC_1 + C_2)\,\delta t .
\]

We now bound $\|\bSigma^{\dagger}-\bSigma^{\ast}\|_{\infty}$.
We define 
\begin{align*}
\bdelta_{\mathcal{I}}^{\ast}(\bz_t, \bz_{t+\delta, \mathcal{I}}) &= \Delta \bz_{t,\mathcal{I}} - \mathbb{E}_{\bx_{t+\delta t}\mid \bx_t} [ \Delta \bz_{t,\mathcal{I}}]\\
&= \Delta \bz_{t,\mathcal{I}} - \mathbb{E}_{\hat{q}} [ \Delta \bz_{t,\mathcal{I}}] 
\end{align*}
Since the encoder is uniformly bounded by $M$, we have: 
\begin{align*}
&\|\bdelta^{\ast}_{\mathcal I}\|_{\infty}
\le 4M, \\ 
&\|\bdelta^{\ast}_{\mathcal I}(\bdelta^{\ast}_{\mathcal I})^{T}\|_{\infty}
\le (4M)^2 = 16M^2 .
\end{align*}
By making use of the following decomposition of $\bdelta^{\ast}$: 
\begin{equation*}
\begin{aligned}    
    \bdelta^{\ast} &= \Delta \bz_t - \bmu^{\ast} (\bz_t)\delta t \\
     &= \textstyle \frac{1}{K}\sum_{\mathcal{I}} \bdelta_{\mathcal{I}}^{\ast} + (\Delta \bz_t - \frac{1}{K}\sum_{\mathcal{I}}\Delta \bz_{t, \mathcal{I}})  \\
     &\textstyle- \mathbb{E}_{\bx_{t+\delta t}\mid \bx_t}[\Delta \bz_t - \frac{1}{K}\sum_{\mathcal{I}}\Delta \bz_{t, \mathcal{I}}]
\end{aligned}
\end{equation*}
we have 
\begin{equation*}
\begin{aligned}
     &\quad \lVert\bSigma^{\dagger}(\bz_t) - \bSigma^{\ast}(\bz_t) \rVert \\
     &= \textstyle \lVert \frac{1}{K\delta t}\mathbb{E}_{\mathcal{I}}\mathbb{E}_q[\bdelta_{\mathcal{I}}^{\dagger} (\bdelta_{\mathcal{I}}^{\dagger})^T] - \frac{1}{\delta t}\mathbb{E}_{\bx_{t+\delta t\mid \bx_t}} [\bdelta^{\ast} (\bdelta^{\ast})^T ] \rVert  \\
     &\leq \textstyle \lVert \frac{1}{K\delta t}\mathbb{E}_{\mathcal{I}}\mathbb{E}_q[\bdelta_{\mathcal{I}}^{\dagger} (\bdelta_{\mathcal{I}}^{\dagger})^T] - \frac{1}{K^2\delta t}\mathbb{E}_{\bx_{t+\delta t\mid \bx_t}} [\sum_{\mathcal{I}, \mathcal{J}}\bdelta^{\ast}_{\mathcal{I}} (\bdelta^{\ast}_{\mathcal{J}})^T ] \rVert \\
     &\textstyle + 4 \lVert  \frac{1}{K}\sum_{\mathcal{I}} \bdelta_{\mathcal{I}}^{\ast} \rVert_{\infty} \cdot \lVert\mathbb{E}_{\bx_{t+\delta t} \mid \bx_t } [  \Delta\bz_t - \frac{1}{K}\sum_{\mathcal{I}}\Delta \bz_{t, \mathcal{I}} ] \rVert \\
     &+\mathcal{O}(\delta t^3) \\ 
     &\leq \textstyle \lVert \frac{1}{K\delta t}\mathbb{E}_{\mathcal{I}}\mathbb{E}_q[\bdelta_{\mathcal{I}}^{\dagger} (\bdelta_{\mathcal{I}}^{\dagger})^T] - \frac{1}{K\delta t}\mathbb{E}_{\mathcal{I}}\mathbb{E}_{\bx_{t+\delta t\mid \bx_t}} [\bdelta^{\ast}_{\mathcal{I}} (\bdelta^{\ast}_{\mathcal{I}})^T ] \rVert \\
     &+ \lVert \textstyle \frac{1}{K^2\delta t} \mathbb{E}_{\bx_{t+\delta t}\mid \bx_t} [\sum_{\mathcal{I}\neq \mathcal{J}}  \bdelta^{\ast}_{\mathcal{I}} (\bdelta^{\ast}_{\mathcal{J}})^T] \rVert \\
     &+ 16MC_2\delta t +\mathcal{O}(\delta t^3) \\ 
\end{aligned}
\end{equation*}

Using the decomposition $\bdelta_\mathcal{I}^\ast = \bdelta_\mathcal{I}^{\dagger} + (\bmu^{\dagger} - \mathbb{E}_{\hat{q}}[\Delta \bz_{t, \mathcal{I}}] )$, we can decompose $\bdelta_{\mathcal{I}}^{\ast} (\bdelta_{\mathcal{I}}^{\ast})^T$:
\begin{align*}
    \bdelta_{\mathcal{I}}^{\ast} (\bdelta_{\mathcal{I}}^{\ast})^T &=  \bdelta_{\mathcal{I}}^{\dagger} (\bdelta_{\mathcal{I}}^{\dagger})^T \\
    &+ \bdelta_{\mathcal{I}}^{\dagger} (\bmu^{\dagger} - \mathbb{E}_{\hat{q}}[\Delta \bz_{t, \mathcal{I}}] )^T   + (\bmu^{\dagger} - \mathbb{E}_{\hat{q}}[\Delta \bz_{t, \mathcal{I}}] ) (\bdelta_{\mathcal{I}}^{\dagger})^T \\
    &+  (\bmu^{\dagger} - \mathbb{E}_{\hat{q}}[\Delta \bz_{t, \mathcal{I}}] )(\bmu^{\dagger} - \mathbb{E}_{\hat{q}}[\Delta \bz_{t, \mathcal{I}}] ) ^T
\end{align*}
We have  
\begin{equation*}
\begin{aligned}
    &\quad \textstyle \lVert \frac{1}{K\delta t}\mathbb{E}_{\mathcal{I}}\mathbb{E}_q[\bdelta_{\mathcal{I}}^{\dagger} (\bdelta_{\mathcal{I}}^{\dagger})^T] - \frac{1}{K\delta t}\mathbb{E}_{\mathcal{I}}\mathbb{E}_{\bx_{t+\delta t\mid \bx_t}} [\bdelta^{\ast}_{\mathcal{I}} (\bdelta^{\ast}_{\mathcal{I}})^T ] \rVert  \\
    &\textstyle \leq \tfrac{1}{K\delta t} \lVert \mathbb{E}_{\mathcal{I}} \mathbb{E}_{q} [ \bdelta_\mathcal{I}^{\dagger} (\bdelta_\mathcal{I}^{\dagger})^T ]  - \mathbb{E}_{\mathcal{I}} \mathbb{E}_{\hat{q}} [ \bdelta_\mathcal{I}^{\dagger} (\bdelta_\mathcal{I}^{\dagger})^T ]  \rVert \\
    &\textstyle + \frac{2}{K\delta t}\lVert \mathbb{E}_{\mathcal{I}} \mathbb{E}_{\hat{q}}  [\bdelta_{\mathcal{I}}^{\dagger} (\bmu^{\dagger} - \mathbb{E}_{\hat{q}}[\Delta \bz_{t, \mathcal{I}}] )^T ] \rVert \\
     &\textstyle + \frac{1}{K\delta t}\lVert\mathbb{E}_{\mathcal{I}}  [(\bmu^{\dagger} - \mathbb{E}_{\hat{q}}[\Delta \bz_{t, \mathcal{I}}] )(\bmu^{\dagger} - \mathbb{E}_{\hat{q}}[\Delta \bz_{t, \mathcal{I}}] ) ^T] \rVert
\end{aligned}
\end{equation*}
We can bound the first term by 
\begin{align*}
    &\quad \textstyle \tfrac{1}{K\delta t} \lVert \mathbb{E}_{\mathcal{I}} \mathbb{E}_{q} [ \bdelta_\mathcal{I}^{\dagger} (\bdelta_\mathcal{I}^{\dagger})^T ]  - \mathbb{E}_{\mathcal{I}} \mathbb{E}_{\hat{q}} [ \bdelta_\mathcal{I}^{\dagger} (\bdelta_\mathcal{I}^{\dagger})^T ]  \rVert \\
    &\leq\textstyle \frac{2 \cdot (4M)^2}{K \delta t} \delta_{\text{TV}}(q, \hat{q}) \\
    &\leq \textstyle\frac{32M^2C_1 }{K} \delta t 
\end{align*}
Next, we bound the second term  
\begin{align*}
    &\textstyle\frac{2}{K\delta t}\lVert \mathbb{E}_{\mathcal{I}} \mathbb{E}_{\hat{q}}  [\bdelta_{\mathcal{I}}^{\dagger} (\bmu^{\dagger} - \mathbb{E}_{\hat{q}}[\Delta \bz_{t, \mathcal{I}}] )^T ] \rVert \\
    &\leq\textstyle \frac{2}{K\delta t} \lVert \bdelta_{\mathcal{I}}^{\dagger} \rVert_{\infty} \cdot \lVert \mathbb{E}_{\mathcal{I}}\mathbb{E}_{\hat{q}}[\Delta \bz_{t, \mathcal{I}}] - \mathbb{E}_{\mathcal{I}}\mathbb{E}_{q}[\Delta \bz_{t, \mathcal{I}}]  \rVert \\
    &\leq \textstyle \frac{8M}{K\delta t} \cdot 4M \cdot \delta_{\text{TV}}(q, \hat{q}) \\
    &\textstyle \leq \frac{32M^2C_1}{K}\delta t
\end{align*}
We can also bound the third term  
\begin{align*}
    & \textstyle  \frac{1}{K \delta t}\lVert\mathbb{E}_{\mathcal{I}}  [(\bmu^{\dagger} - \mathbb{E}_{\hat{q}}[\Delta \bz_{t, \mathcal{I}}] )(\bmu^{\dagger} - \mathbb{E}_{\hat{q}}[\Delta \bz_{t, \mathcal{I}}] ) ^T] \rVert \\
    &\leq \textstyle \frac{1}{K \delta t}\lVert \mathbb{E}_{\mathcal{I}}\mathbb{E}_{\hat{q}}[\Delta \bz_{t, \mathcal{I}}] - \mathbb{E}_{\mathcal{I}}\mathbb{E}_{q}[\Delta \bz_{t, \mathcal{I}}]  \rVert_{\infty}^2 \\
    &\textstyle \leq \frac{1}{K \delta t} \delta_{\text{TV}}(q, \hat{q})^2 \\
    &\textstyle \leq \frac{C_1^2}{K} \delta t^3
\end{align*}
Combining the above bounds, we conclude that
\begin{align*}
    &\quad \textstyle \lVert \frac{1}{K\delta t}\mathbb{E}_{\mathcal{I}}\mathbb{E}_q[\bdelta_{\mathcal{I}}^{\dagger} (\bdelta_{\mathcal{I}}^{\dagger})^T] - \frac{1}{K\delta t}\mathbb{E}_{\mathcal{I}}\mathbb{E}_{\bx_{t+\delta t\mid \bx_t}} [\bdelta^{\ast}_{\mathcal{I}} (\bdelta^{\ast}_{\mathcal{I}})^T ] \rVert  \\ 
    &\textstyle\leq \frac{64M^2 C_1}{K} \delta t + \mathcal{O}(\delta t^3)
\end{align*}

By the assumption that increments $\bz_{t+\delta t, \mathcal{I}} - \bz_t$ and $\bz_{t+\delta t, \mathcal{J}} - \bz_t$  are independent for disjoint $\mathcal{I} \neq \mathcal{J}$ given $\bx_t$, the cross-covariance terms vanish. Specifically, 
\begin{equation*}
    \mathbb{E}_{\bx_{t+\delta t}|\bx_t} \left[\bdelta_{\mathcal{I}}^{\ast} (\bdelta_{\mathcal{J}}^{\ast})^T \right] = \mathbb{E}_{\bx_{t+\delta t}|\bx_t} \left[\bdelta_{\mathcal{I}}^{\ast}\right] \mathbb{E}_{\bx_{t+\delta t}|\bx_t} \left[(\bdelta_{\mathcal{J}}^{\ast})^T \right]
\end{equation*}
Since
\begin{equation*}
    \mathbb{E}_{\bx_{t+\delta t}|\bx_t} \left[\bdelta_{\mathcal{I}}^{\ast}\right] = \mathbb{E}_{\bx_{t+\delta t}|\bx_t}[\Delta \bz_{t, \mathcal{I}}] - \mathbb{E}_{\bx_{t+\delta t}|\bx_t}[\Delta \bz_{t, \mathcal{I}}] = 0
\end{equation*}
we have  
\begin{equation*}
    \mathbb{E}_{\bx_{t+\delta t}|\bx_t} \left[\bdelta_{\mathcal{I}}^{\ast} (\bdelta_{\mathcal{J}}^{\ast})^T \right] = 0 
\end{equation*}
which yields 
\begin{equation*}
\begin{aligned}
     \quad \lVert\bSigma^{\dagger}(\bz_t) - \bSigma^{\ast}(\bz_t) \rVert 
     \textstyle\leq \frac{64M^2 C_1}{K} \delta t +  16MC_2\delta t+  \mathcal{O}(\delta t^3)
\end{aligned}
\end{equation*}
\end{proof}

\section{Additional Experimental Results}\label{CW}
\subsection{Curie--Weiss model}
\begin{figure}[!t]
    \centering
    \includegraphics[width=0.6\linewidth]{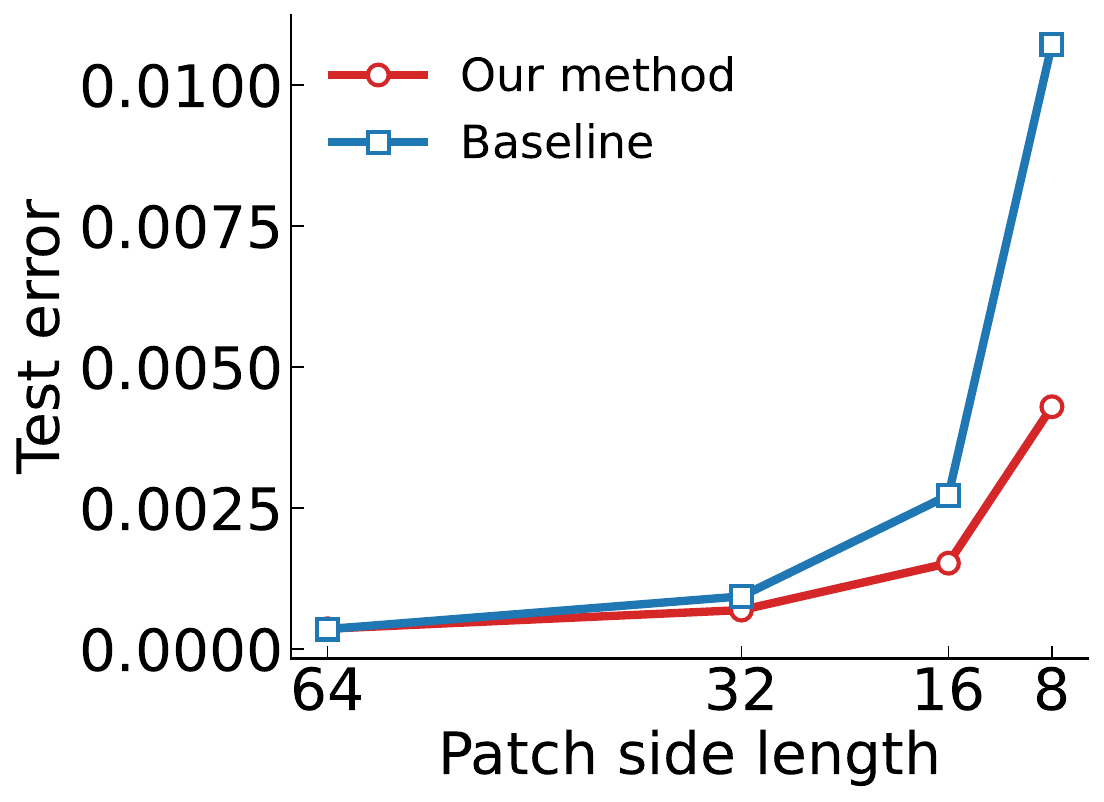}
    \caption{Results on the Curie--Weiss model, where the test error is plotted as a function of $n_s$. The test error is the mean relative error of the mean macroscopic observables between ground-truth and predicted trajectories.}
    \label{fig:CW_scale}
\end{figure}
\begin{figure}[!t]
    \centering
    \includegraphics[width=1.0\linewidth]{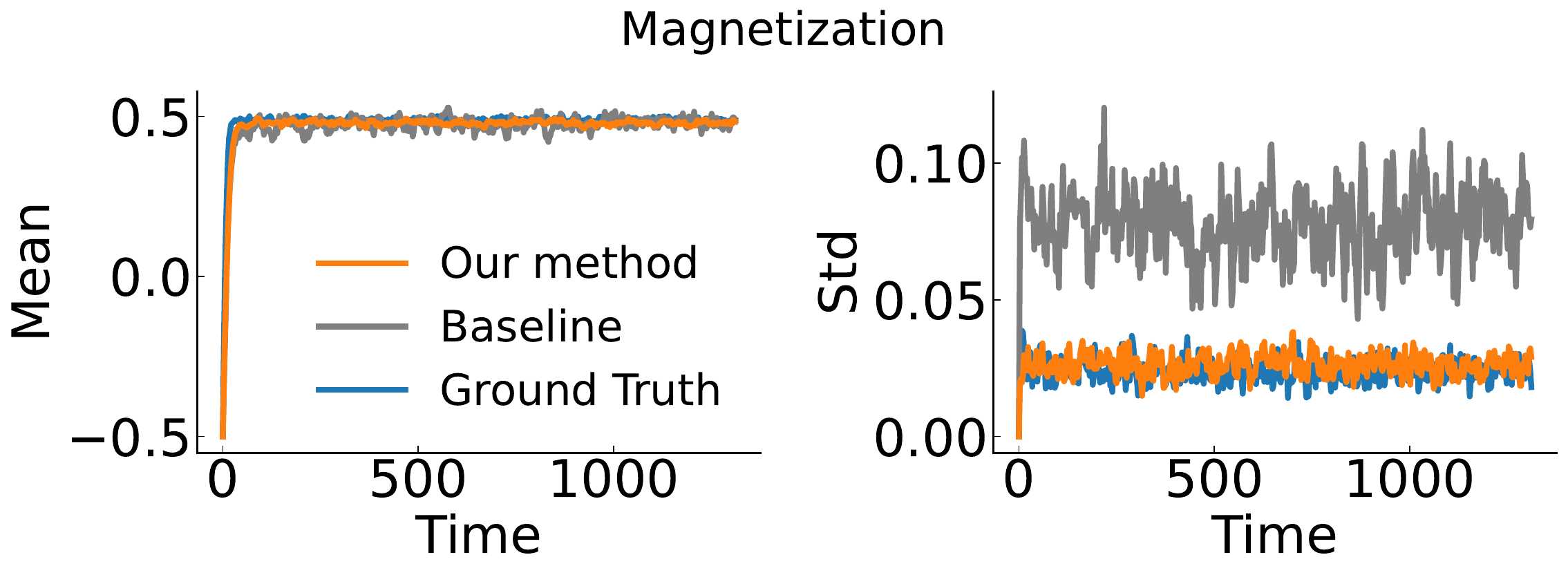}
    \caption{Results on the Curie--Weiss model with $n_s=16^2$. Mean and standard deviation of the magnetization are estimated from $20$ trajectories per method. 
    }
    \label{fig:CW_traj}
\end{figure}
As an additional robustness check, we consider the Curie--Weiss model, which does not strictly satisfy the assumptions of our framework.
The Curie--Weiss model is a mean-field system with nonlocal interactions~\cite{kochmanski2013curie, ellis2012entropy}.
Each spin interacts with all others through a global coupling. 
The Hamiltonian of the Curie--Weiss model with $n$ spins is given by: 
\begin{equation*}\label{eq:CW-energy}
    H_{n, h}(\sigma) = -\frac{J}{2n} \sum_{i,j=1}^n \sigma_i \sigma_j - h \sum_{i=1}^n \sigma_i  ,
\end{equation*}
where $\sigma_i \in \{-1,1\}$ and $h$ denotes the external magnetic field. 
The Hamiltonian can also be expressed as a function of the magnetization $M = \sum_i \sigma_i / n$:
\begin{equation*}
    H_{n, h}(\sigma) = -n\left[ \frac{J}{2} \left(\frac{\sum_i \sigma_i}{n}\right)^2 + h \frac{\sum_i \sigma_i}{n} \right] .
\end{equation*}
Hence, the microscopic dynamics can be fully characterized by the magnetization. We choose the magnetization as the macroscopic observable of interest since the magnetization is closed by itself, and no additional closure variables are needed. As in the Ising model, we adopt the continuous-time Glauber dynamics as the microscopic dynamics.

The Curie--Weiss model does not fit directly into our framework, because it is a mean-field model without any explicit spatial structure. To handle this, we imagine the spins are arranged on a square lattice.  

We repeat the ablation study described in \cref{exp:ising} with $h=0.1$ and $T=1.1 > T_c=1$, using the same \textsc{Upsample} and \textsc{LocalRelax} steps.
\cref{fig:CW_scale} and \cref{fig:CW_traj} report the corresponding results.
We observe that our method consistently outperforms the baseline when $n_s< n$.
Moreover, the trajectory statistics produced by our method closely match those of the ground truth, whereas the baseline exhibits much larger variability.
Although the Curie--Weiss model does not strictly satisfy the locality assumptions underlying our framework, our method continues to yield reasonable macroscopic predictions. These observations suggest empirical robustness of our framework.

\subsection{Linear Chain with boundary driving}
\begin{table*}[!ht]
\centering
\begin{tabular}{l|cc|cc}
\hline
Parameter & \multicolumn{2}{c|}{Small System ($N=10$)} & \multicolumn{2}{c}{Large System ($KN=100$)} \\
 & Theoretical & Learned (Conventional) & Theoretical & Learned (Ours) \\ \hline
Slope $a$ ($-\gamma$) & $-0.10$ & $-0.1047$ & $-0.10$ & $-0.0929$ \\
Bias $b$ ($F/N$) & $\mathbf{1.50}$ & $\mathbf{1.5002}$ & $\mathbf{0.15}$ & $\mathbf{0.1419}$ \\
Diffusion $c$ ($\sigma/\sqrt{N}$) & $0.316$ & $0.3122$ & $0.100$ & $0.1000$ \\ \hline
\end{tabular}
\caption{Comparison of theoretical and learned parameters for the reduced macroscopic SDE}
\label{tab:comparison}
\end{table*}
Our method aims to learn the macroscopic dynamics of a large system using only data generated from small-system simulations. Consider a small system with $N$ particles and a large system with $KN$ particles governed by the same microscopic dynamics. In the modified SDE loss, we multiply the noise covariance by a factor $K$, which is equivalent to scaling the diffusion coefficient by $1/\sqrt{K}$.

A natural question then arises: instead of modifying the loss, could one simply learn the macroscopic dynamics of the small system using the conventional method, and then rescale the diffusion term by $1/\sqrt{K}$ at prediction time to obtain the macroscopic dynamics of the large system? The answer is, in general, no.

The reason is that, for a given macroscopic observable, increasing the system size may change not only the diffusion level but also the macroscopic drift. We illustrate this point using an analytical stochastic differential equation (SDE) example.

Specifically, to demonstrate why training on a small system and merely rescaling the diffusion is insufficient, we consider a one-dimensional chain of $N$ particles with linear on-site friction $\gamma>0$ and nearest-neighbor coupling $\kappa>0$. The system is driven by a constant external force $F$ applied only to the first particle. The microscopic dynamics for the displacement $X_i(t)$ are given by
\begin{equation*}
\begin{aligned}
dX_1
&= \big[-\gamma X_1 + \kappa(X_2 - X_1) + F\big]\,dt
  + \sigma\, dW_1, \\
dX_i
&= \big[-\gamma X_i + \kappa(X_{i-1} - 2X_i + X_{i+1})\big]\,dt + \sigma\, dW_i, \\
&\quad \qquad i = 2,\dots,N-1, \\
dX_N
&= \big[-\gamma X_N + \kappa(X_{N-1} - X_N)\big]\,dt
  + \sigma\, dW_N .
\end{aligned}
\end{equation*}
where $\{W_i\}_{i=1}^N$ are independent Brownian motions.

Let the macroscopic observable be the mean displacement $m_N(t)=\frac{1}{N}\sum_{i=1}^N X_i(t)$. 
Summing the microscopic equations over $i$ yields the closed macroscopic dynamics:
\begin{equation*}
\begin{aligned}
\label{eq:macro_driven_exact}
dm_N(t) &= \underbrace{\left[ -\gamma\, m_N(t) + \frac{F}{N} \right]}_{\text{drift } f_N(m)} dt
+ \underbrace{\frac{\sigma}{\sqrt{N}}}_{\text{diffusion } g_N}\, dB(t), \\
dB(t)&=\frac{1}{\sqrt{N}}\sum_{i=1}^N dW_i(t).
\end{aligned}
\end{equation*}

Now consider a small system of size $N$ and a large system of size $KN$, where the same boundary driving is applied to a single particle. Even though the diffusion term rescales as $1/\sqrt{K}$, the drift terms are also different: 
\[
f_{KN}(m) = -\gamma m + \frac{F}{KN}
\neq
-\gamma m + \frac{F}{N} = f_N(m).
\]
This shows that learning from a small system and only adjusting the diffusion is insufficient to recover the correct macroscopic drift in the large system.

We also demonstrate the difference in the learned drift through experiments. 
We set $N=10$, $K=10$, $F=15$, $\sigma=1$, $\gamma=0.1$, and $\kappa=1$. 
We compare two training strategies:
\begin{enumerate}[label=(\roman*)]
    \item Conventional method. 
    We simulate a small system of size $N$ to obtain pairs $(x_t, x_{t+\delta t})$ from the full microscopic dynamics, and train an SDE model using the standard loss $\mathcal{L}$.
    \item Our method. 
    We simulate a system of size $KN$, partition it into $K$ local patches, and generate partial updates $\bx_{t+\delta t,\mathcal{I}}$ using our partial evolution scheme. 
    We then train the macroscopic SDE using our modified loss $\mathcal{L}_p$.
\end{enumerate}

Since the true macroscopic drift is linear and the diffusion is constant, we restrict the drift network to be linear and the diffusion network to output a constant. 
Specifically, we fit
\[
dm(t) = (a\, m(t) + b)\, dt + c\, dB(t),
\]
so that each method estimates only three parameters $(a,b,c)$. 
The results are summarized in Table~\ref{tab:comparison}. 
We observe that the conventional method accurately recovers the macroscopic dynamics of the small system, whereas our method accurately recovers the macroscopic dynamics of the large system.

\subsection{Additional experiment of Ising model}
We next validate the above observation using the two-dimensional Ising model from the main paper. We take the small system to be a $16\times16$ Ising model and the large system to be $64\times64$. We consider temperature $T=2.25$ and external field $h=0$, where the temperature is close to the critical temperature. 

\begin{figure*}[ht]
    \centering
    \begin{tabular}{@{}c@{\hspace{5pt}}@{\hspace{5pt}}c@{}}
        \quad \ \  {(a)  }& \ \ \ {(b) } \\
         \includegraphics[width=0.25\textwidth]{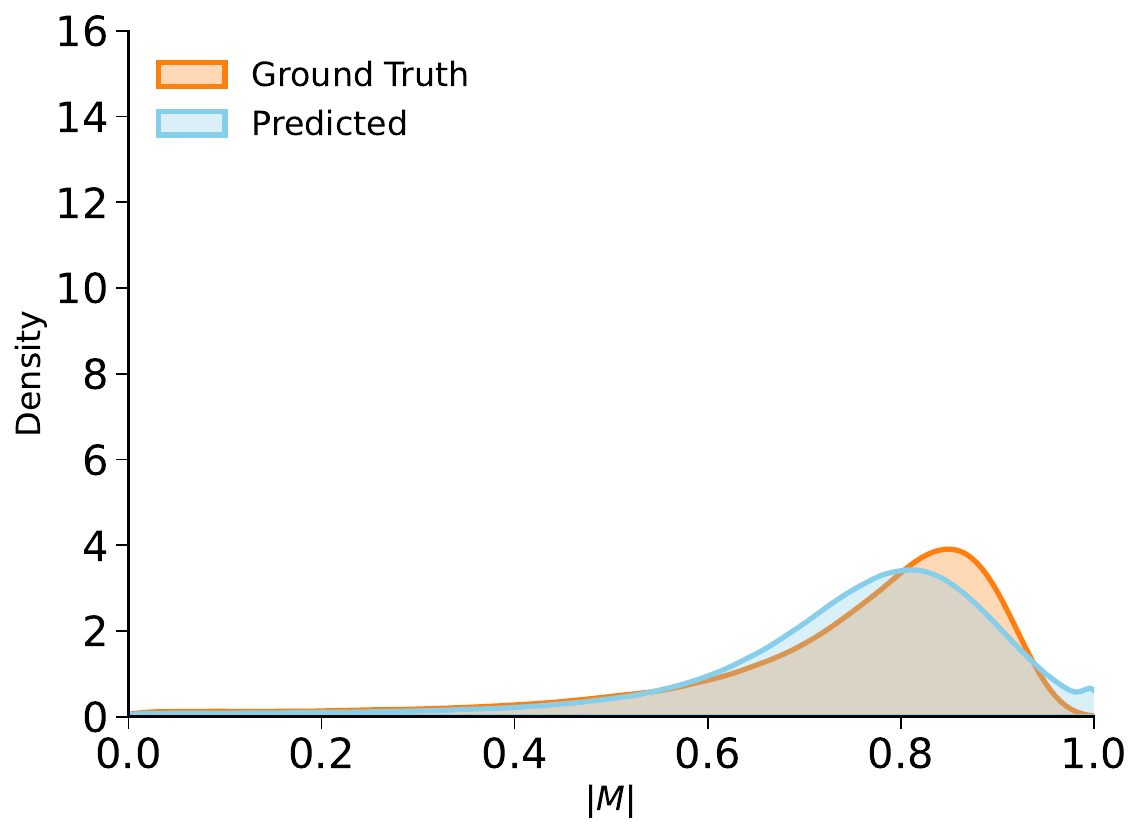} & \includegraphics[width=0.72\textwidth]{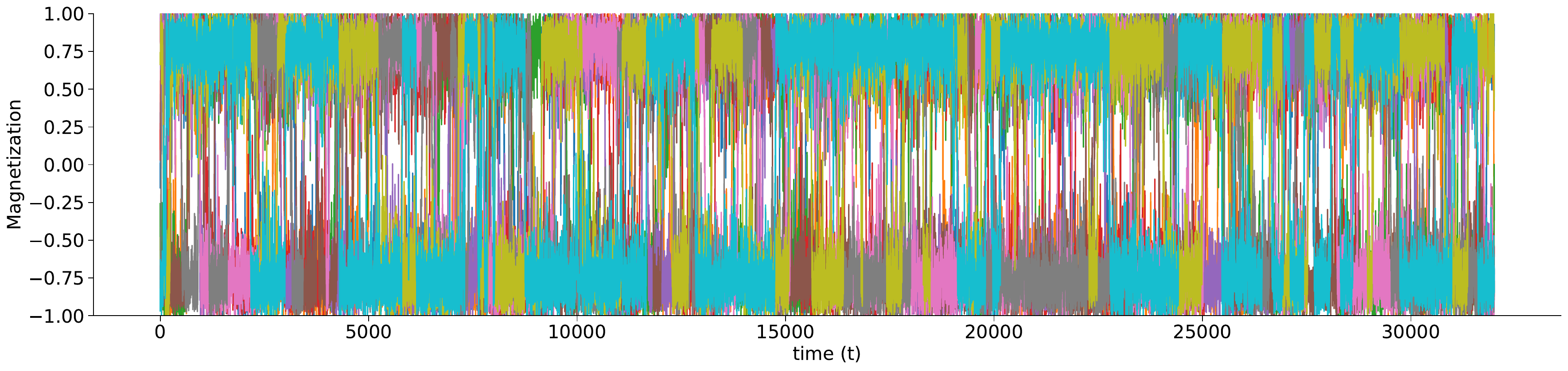} \vspace{-2pt} \\
         \quad \ \  {(c)  } & \ \ \ {(d) } \\
         \includegraphics[width=0.25\textwidth]{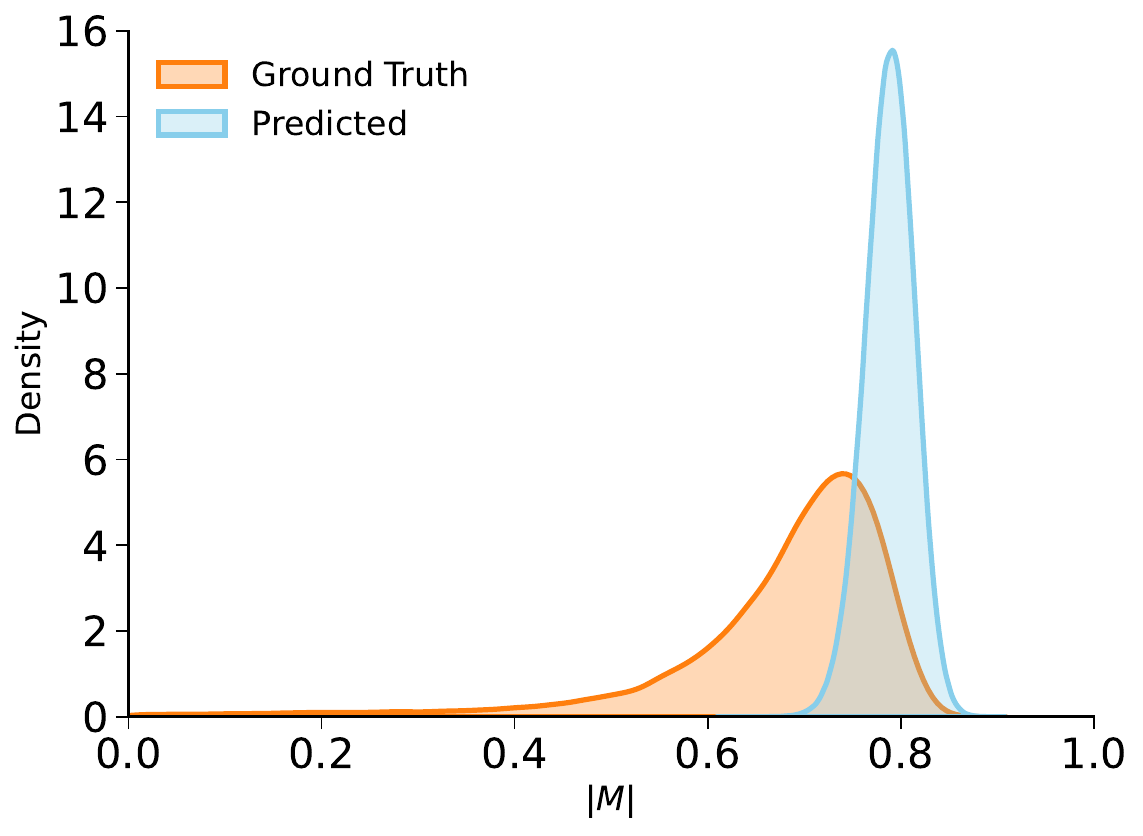}  &
         \includegraphics[width=0.72\textwidth]{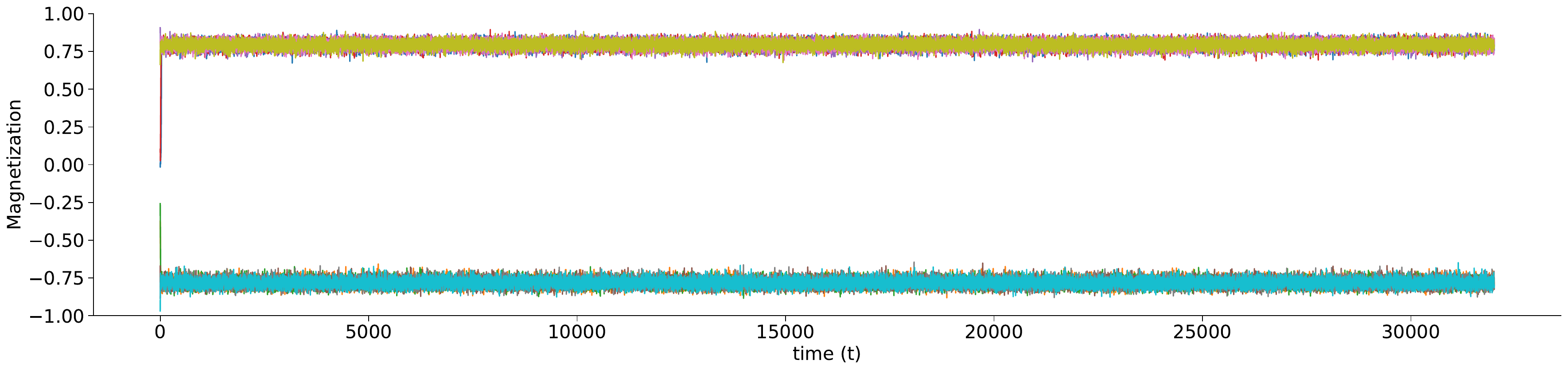} \vspace{-2pt} \\
         \quad \ \  {(e)  } & \ \ \ {(f) } \\
         \includegraphics[width=0.25\textwidth]{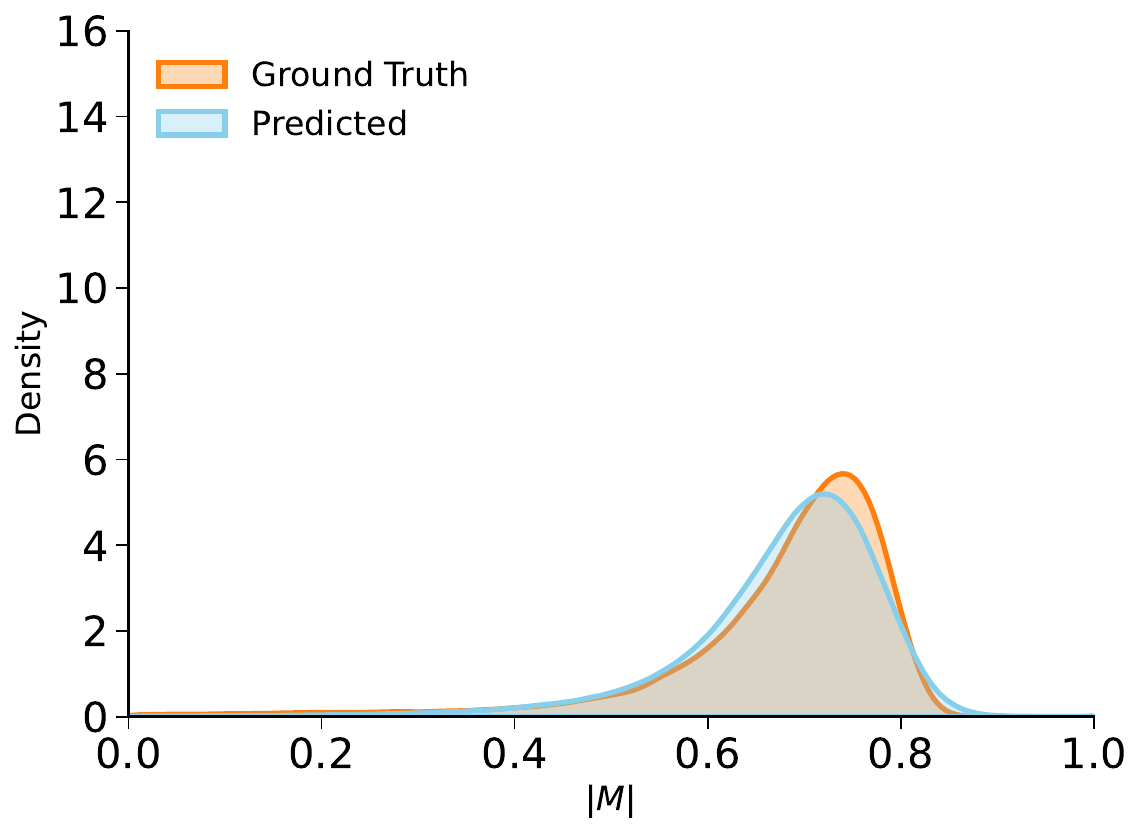}  &
         \includegraphics[width=0.72\textwidth]{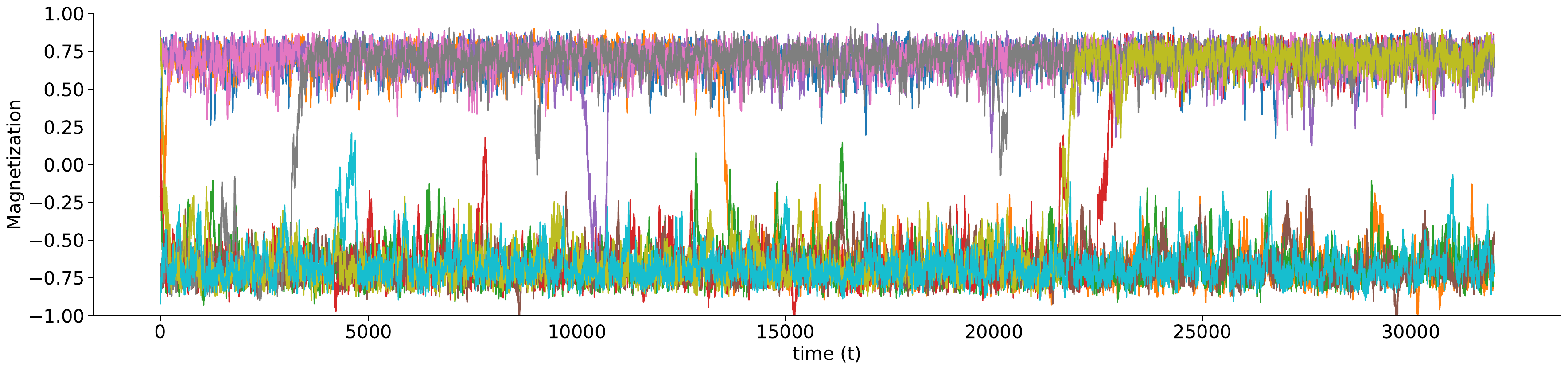} \vspace{-2pt} \\
    \end{tabular}
    \caption{Comparison of magnetization statistics and dynamics. Each row corresponds to settings (i)--(iii) from top to bottom. The first column shows the predicted equilibrium magnetization distribution together with the ground truth. In the first row, the ground truth is obtained from direct simulations of a $16\times16$ Ising system, while in the second and third rows it is obtained from direct simulations of a $64\times64$ system. The second column shows representative predicted trajectories of the magnetization dynamics.}
    \label{fig:compare}
\end{figure*}

We compare the following three settings:
\begin{enumerate}[label=(\roman*)]
    \item $\bx_t$ is sampled from snapshots of small-system trajectories, and $\bx_{t+\delta t}$ is obtained by fully evolving the small system for a short time. The model is trained with the conventional SDE loss $\mathcal{L}$.
    \item We use the same SDE as in (i). We scale the diffusion term of the SDE by $1/\sqrt{16}=1/4$ during prediction. 

    \item $\bx_t$ is sampled from snapshots of large-system trajectories, and $\bx_{t+\delta t,\mathcal{I}}$ is generated by the partial evolution scheme. The model is trained with our modified SDE loss. 
\end{enumerate}
The results are shown in \cref{fig:compare}. 
Since $T=2.25$ is close to the critical temperature, the Ising model undergoes frequent magnetization reversals, and the magnetization therefore changes sign repeatedly over time.

We observe that the SDE trained under setting (i) accurately reproduces the equilibrium magnetization distribution of the $16\times16$ Ising model. 
However, when the same SDE is used for prediction with the diffusion term scaled by $1/4$ as in setting (ii), the predicted magnetization no longer exhibits magnetization reversals. As a consequence, the resulting equilibrium distribution deviates significantly from that of the $64\times64$ Ising model. In contrast, under setting (iii), our method successfully reproduces the equilibrium magnetization distribution of the $64\times64$ Ising model.

The failure of setting (ii) demonstrates that simply rescaling the diffusion term is insufficient for extrapolating macroscopic dynamics from small to large systems. 
This is due to the difference in the drift terms between the small-system macroscopic dynamics and the large-system macroscopic dynamics. 
This shows that learning macroscopic dynamics from small-system simulations alone is nontrivial, highlighting the necessity and importance of our method.

\subsection{Ablation Study}

The conventional SDE loss $\mathcal{L}$ in \cref{eq:L} and our modified SDE loss $\mathcal{L}_p$ in \cref{eq:L_p} involve two data distributions: the distribution of input states $\mathbf{x}_t$ and the distribution of short-time evolved states $\mathbf{x}_{t+\delta t}$ or $\mathbf{x}_{t+\delta t,\mathcal{I}}$. Accordingly, when relying on small-system simulations, two corresponding sources of distribution shift arise. 

The first source arises from a mismatch in the distribution of $\mathbf{x}_t$, as configurations constructed from small-system simulations generally differ from snapshots sampled from large-system trajectories. To mitigate this effect, we introduce a hierarchical upsampling scheme to construct large-system configurations from small-system snapshots.

The second source of distribution shift concerns the distribution of the short-time evolved states. In particular, the partial evolution scheme induces a mismatch between the distribution of $\mathbf{x}_{t+\delta t,\mathcal I}$ and that of $\mathbf{x}_{t+\delta t}$ obtained by evolving the full microscopic system. To address this effect, we introduce the modified SDE loss $\mathcal{L}_p$ together with Theorem~1, which provides a theoretical justification for correcting the statistical bias introduced by partial evolution. Importantly, Theorem~1 applies specifically to the distribution shift in $\mathbf{x}_{t+\delta t,\mathcal I}$, while the distribution shift in $\mathbf{x}_t$ is shared by both the conventional loss $\mathcal{L}$ and the modified loss $\mathcal{L}_p$. This learning formulation is general and applicable across different systems.

To disentangle these two sources of distribution shift, we perform an ablation study. 
We consider three ways of constructing the input dataset $\mathbf{x}_t$: 
(i) snapshots sampled from large-system trajectories; 
(ii) snapshots obtained by naive upsampling from small-system trajectories without \textsc{LocalRelax}; and 
(iii) snapshots generated by the hierarchical upsampling scheme.

Correspondingly, we consider three choices for generating the short-time evolved states and the associated training loss: 
(i) Conventional method: full microscopic evolution with the conventional loss $\mathcal{L}$; 
(ii) Baseline: partial evolution with the conventional loss $\mathcal{L}$ ; and 
(iii) Our method: partial evolution with the proposed loss $\mathcal{L}_p$.

\begin{figure*}[t]
    \centering
    \begin{tabular}{@{}c@{\hspace{5pt}}@{\hspace{5pt}}c@{\hspace{2pt}}c@{}}
        \quad \ \  {(a)  }
         & \ \ \
         {(b) }
         &  \ \ 
         {(c) }\\
         \includegraphics[width=0.3\textwidth]{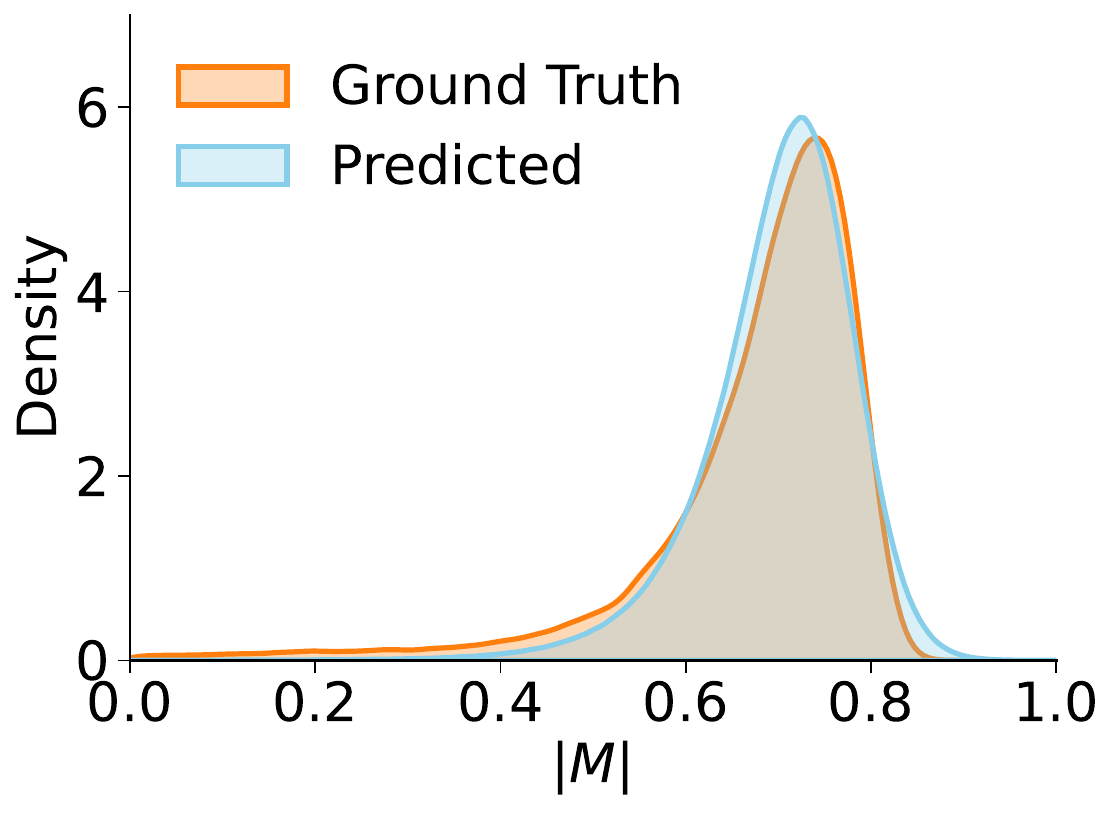} & \includegraphics[width=0.3\textwidth]{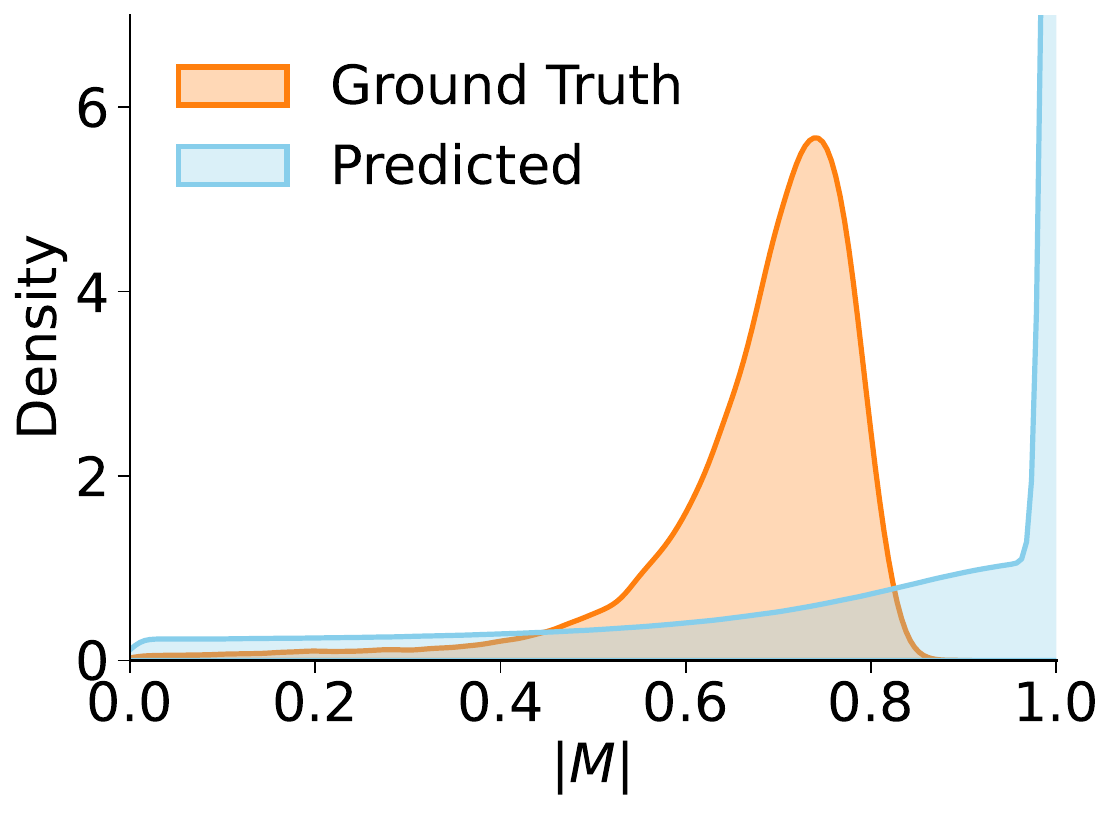} & \includegraphics[width=0.3\textwidth]{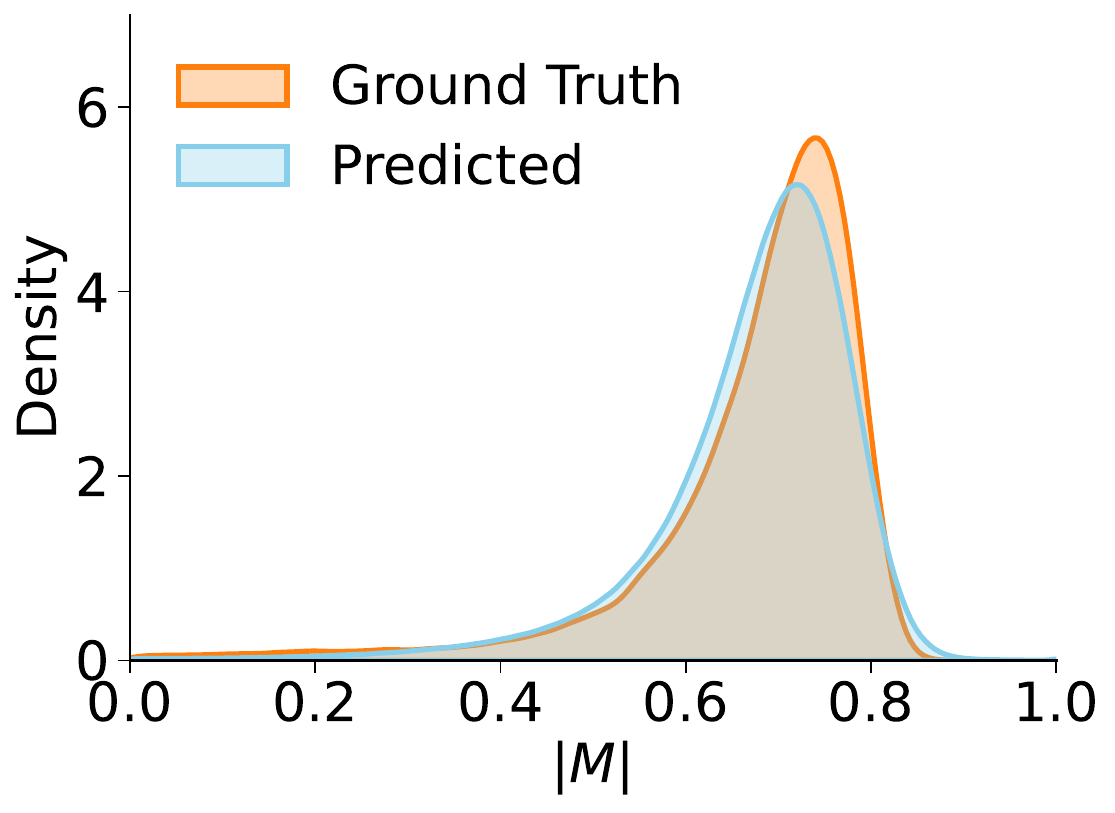}
         \vspace{-5pt}
         \\
         \quad \ \  {(d)  }
         & \ \ \
         {(e) }
         &  \ \ 
         {(f) }\\
         \includegraphics[width=0.3\textwidth]{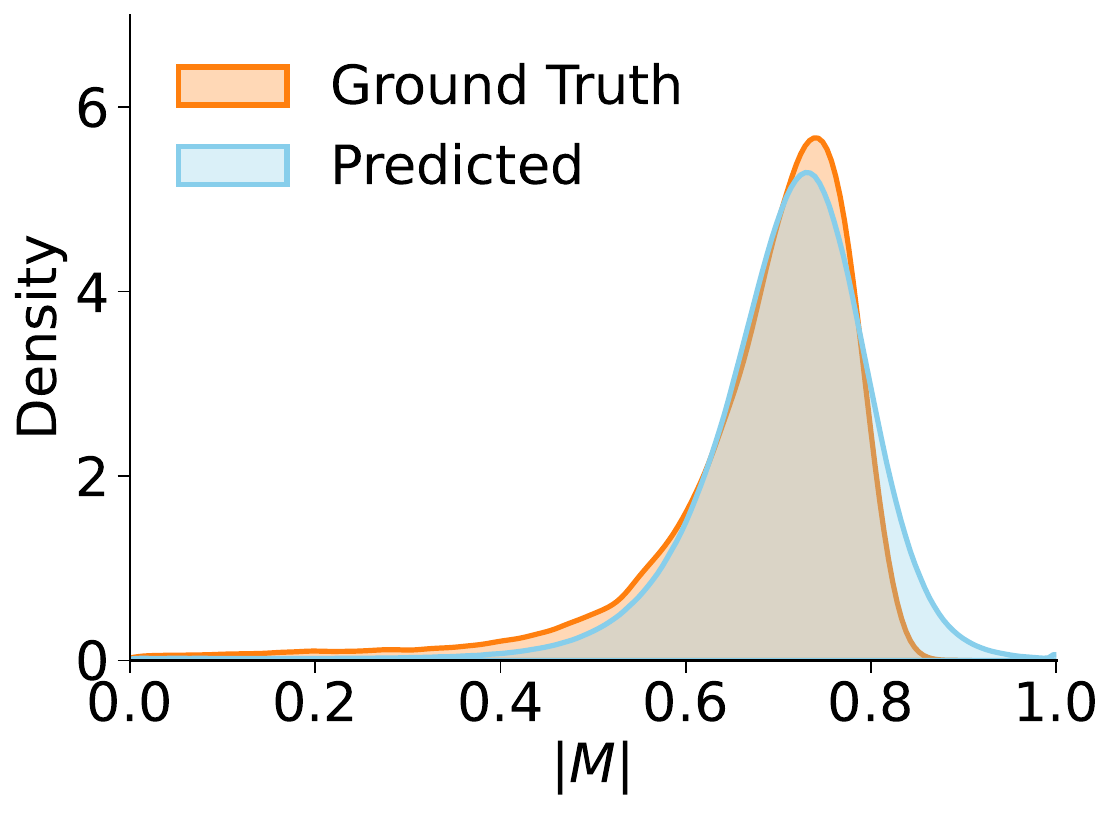}  &
         \includegraphics[width=0.3\textwidth]{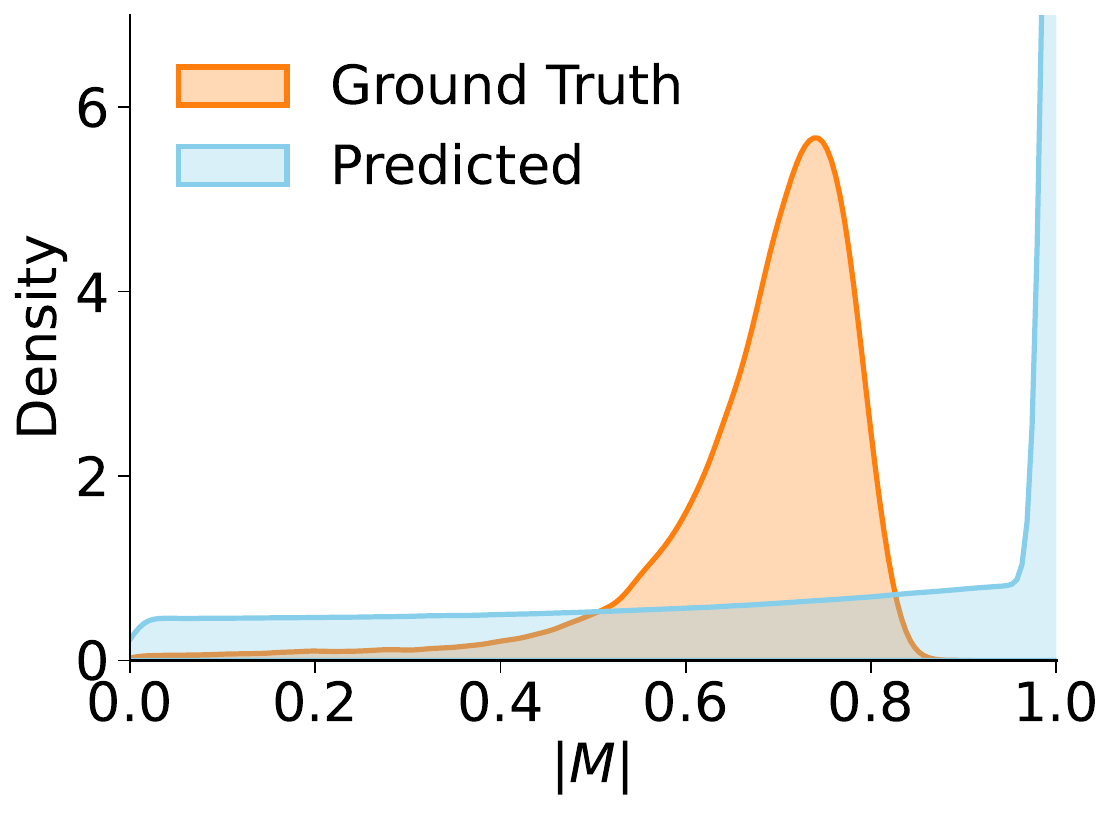} & \includegraphics[width=0.3\textwidth]{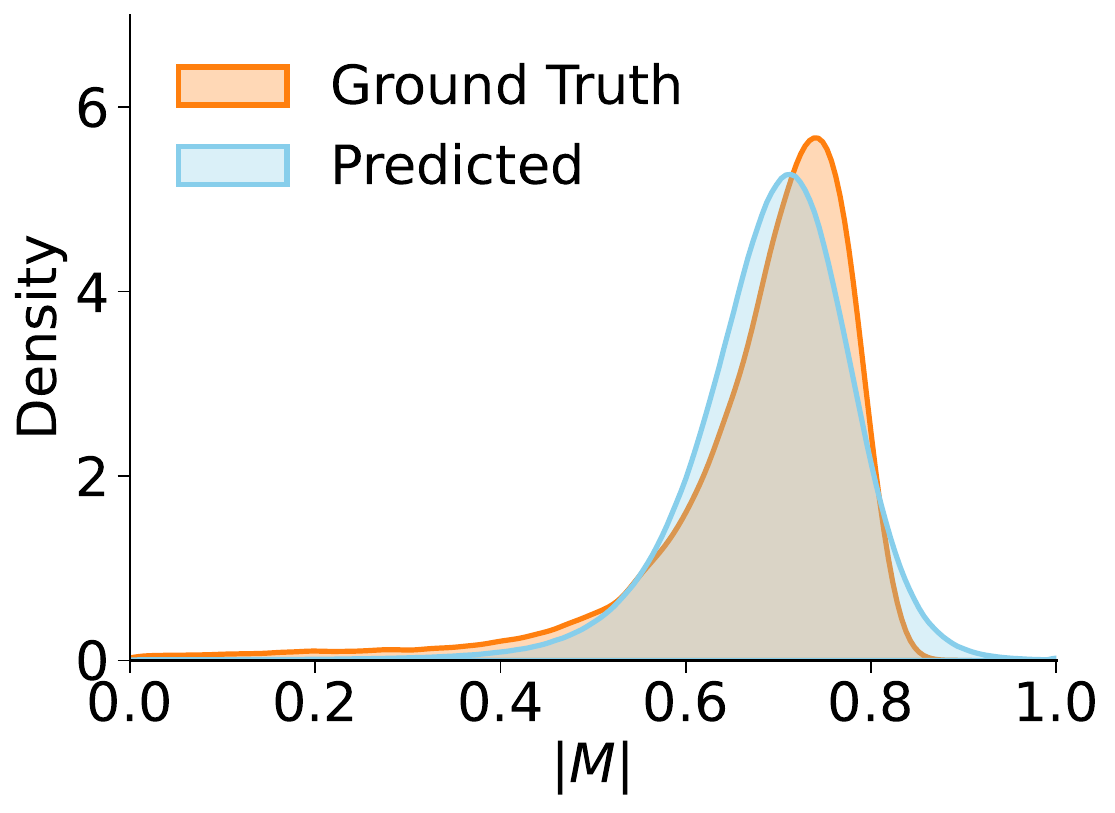} \vspace{-5pt}
         \\
    \end{tabular}
    \caption{
    Equilibrium magnetization distributions obtained by long-term prediction of the learned macroscopic dynamics under different experimental settings.
    Panels (a)--(c) correspond to the first row of \cref{tab:ablation}, where $\mathbf{x}_t$ is sampled from large-system trajectories.
    Panels (d)--(f) correspond to the third row of \cref{tab:ablation}, where $\mathbf{x}_t$ is generated using the hierarchical upsampling scheme.
    From left to right, each column shows results obtained using full microscopic evolution with loss $\mathcal{L}$ (conventional method), partial evolution with loss $\mathcal{L}$ (baseline), and partial evolution with the proposed loss $\mathcal{L}_p$ (our method). 
    }
\label{fig:ablation}
\end{figure*}

The experimental settings are summarized in \cref{tab:ablation}.
The large system is a $64\times64$ Ising model, and the small system is a $16\times16$ Ising model.
We use temperature $T=2.25$ and external field $h=0$.
For each setting, we train the macroscopic dynamics model and perform long-term prediction.
We then compare the predicted equilibrium magnetization distribution with the ground truth.
The corresponding results are shown in \cref{fig:ablation}.

The results for the second row of \cref{tab:ablation} are not shown.
When snapshots obtained by naive upsampling from small-system snapshots are used, the distribution shift in $\mathbf{x}_t$ is too large for the SDE dynamics to be learned. As a result, the long-term predictions diverge and produce NaN values.

\begin{table*}[t]
\centering
\small
\setlength{\tabcolsep}{6pt}
\renewcommand{\arraystretch}{1.35}

\begin{adjustbox}{max width=\linewidth}
\begin{tabular}{l l l l}

\Xhline{3\arrayrulewidth}
Dataset of $\mathbf{x}_t$
& \makecell[l]{
 \textbf{Conventional Method}\\
$\mathbf{x}_{t+\delta t}$: Full evolution\\
Training: loss $\mathcal{L}$
}
& \makecell[l]{
\textbf{Baseline} \\
$\mathbf{x}_{t+\delta t,\mathcal{I}}$: Partial evolution scheme\\
Training: loss $\mathcal{L}$ \\
}
& \makecell[l]{
\textbf{Our method}\\
$\mathbf{x}_{t+\delta t,\mathcal{I}}$: Partial evolution scheme \\
Training loss $\mathcal{L}_p$ 
} \\
\Xhline{1\arrayrulewidth}

$64{\times}64$ Ising trajectory snapshots
& (a) & (b) & (c) \\
\Xhline{1\arrayrulewidth}

\makecell[l]{
\textbf{Upsample (no \textsc{LocalRelax})}\\
Upsampled from $16{\times}16$ snapshots 
} 
& -- & -- & -- \\
\Xhline{1\arrayrulewidth}

\makecell[l]{
\textbf{Hierarchical upsampling scheme}\\
generated from $16{\times}16$ snapshots
} 
& (d) & (e) & (f) \\

\Xhline{3\arrayrulewidth}
\end{tabular}
\end{adjustbox}

\caption{
Summary of experimental settings used in the ablation study. Each table entry corresponds to a specific experimental setting, labeled by a letter that matches the subfigure label in Fig.~\ref{fig:ablation}.
}
\label{tab:ablation}
\end{table*}

From the ablation study, we observe the following:
\begin{enumerate}[label=(\roman*)]
    \item When the distribution shift in $\mathbf{x}_t$ is too large, neither full evolution nor partial evolution can learn the macroscopic dynamics well, regardless of the training loss used.
    \item The \textsc{LocalRelax} step helps remove unphysical artifacts introduced by the \textsc{Upsample} step.
    By comparing the first and second rows of \cref{fig:ablation}, we see that the hierarchical upsampling scheme produces a more reasonable distribution of $\mathbf{x}_t$ than naive upsampling.
    \item Comparing each column, regardless of how $\mathbf{x}_t$ is generated, the proposed loss $\mathcal{L}_p$ consistently mitigates the error introduced by partial evolution compared to the conventional loss.
\end{enumerate}

In summary, the hierarchical upsampling scheme helps produce a more reasonable distribution of $\mathbf{x}_t$ and mitigates the distribution shift in $\bx_t$, while the modified SDE loss acts solely to correct the additional variance introduced by the partial evolution scheme.

\section{Additional Proofs}\label{app:proof}
We will show that the stochastic predator-prey system exactly satisfies the condition of \cref{thm:1}.
\begin{proof}

The spatial domain $[0,1]$ is discretized into $n=200$ uniform grids with
$x_i = \left(i-\tfrac{1}{2}\right)\Delta x$, $\Delta x = 1/200$, $1 \le i \le 200$.
Let
\begin{align*}
    \mathbf{u}_t  &= \big(u(x_1,t),\ldots,u(x_{200},t)\big), \\
\mathbf{v}_t  &= \big(v(x_1,t),\ldots,v(x_{200},t)\big),
\end{align*}
and treat $(\mathbf{u}_t ,\mathbf{v}_t ) \in \mathbb R^{400}$ as the microscopic state. 
We approximate the spatial derivatives using finite differences.
For $2 \le i \le 199$,
\[
\frac{\partial^2 u}{\partial x^2}(x_i,t)
\approx \frac{u(x_{i+1},t) - 2u(x_i,t) + u(x_{i-1},t)}{\Delta x^2},
\]
and at the boundaries,
\begin{align*}
&\frac{\partial^2 u}{\partial x^2}(x_1,t)
\approx \frac{u(x_2,t) - u(x_1,t)}{\Delta x^2}, \\
&\frac{\partial^2 u}{\partial x^2}(x_{200},t)
\approx \frac{u(x_{199},t) - u(x_{200},t)}{\Delta x^2}.
\end{align*}
The same discretization is used for $v$.

Define $h_u(u,v)=u(1-u-v)$ and $h_v(u,v)=av(u-b)$, and let
$h_u(\mathbf u,\mathbf v)$ and $h_v(\mathbf u,\mathbf v)$ denote their
element-wise application. The semi-discrete system can be written as
\begin{align*}
\frac{d\mathbf u}{dt}
&= h_u(\mathbf u,\mathbf v) + c \mathbf A \mathbf u
+ \sigma_u\, d\mathbf B_t^{u}, \\
\frac{d\mathbf v}{dt}
&= h_v(\mathbf u,\mathbf v) + \mathbf A \mathbf v
+ \sigma_v\, d\mathbf B_t^{v},
\end{align*}
where $\mathbf B_t^{u}, \mathbf B_t^{v} \in \mathbb R^{200}$ are independent
standard Brownian motions, and the matrix
$\mathbf A \in \mathbb R^{200 \times 200}$ is given by
\[
\mathbf A
= \frac{1}{\Delta x^2}
\begin{pmatrix}
-1 & 1 & 0 & \cdots & 0 \\
1 & -2 & 1 & \ddots & \vdots \\
0 & 1 & -2 & \ddots & 0 \\
\vdots & \ddots & \ddots & \ddots & 1 \\
0 & \cdots & 0 & 1 & -1
\end{pmatrix}.
\]
For time discretization, we apply the Euler--Maruyama scheme:
\begin{equation*} 
\begin{aligned}
\mathbf u_{t+\delta t}
&= \mathbf u_{t}
+ \delta t \big( h_u(\mathbf u_t,\mathbf v_t)
+ c \mathbf A \mathbf u_t \big)
+ \sigma_u \delta \mathbf B_{t}^{u}, \\
\mathbf v_{t+\delta t}
&= \mathbf v_{t}
+ \delta t \big( h_v(\mathbf u_t,\mathbf v_t)
+ \mathbf A \mathbf v_t \big)
+ \sigma_v \delta \mathbf B_{t}^{v},
\end{aligned}
\end{equation*}
where $\delta \mathbf B_{t}^{u}$ and $\delta \mathbf B_{t}^{v}$ are independent
Gaussian random vectors distributed as
$\mathcal N(\mathbf 0, \delta t \mathbf I_{200})$.

Step 1: $q(\bx_{t+\delta t,\mathcal I} \mid \bx_t) = \hat q(\bx_{t+\delta t,\mathcal I} \mid \bx_t).$
The conditional distribution
of $\bx_{t+\delta t} = (\mathbf u_{t+\delta t}, \mathbf v_{t+\delta t})$
given $\bx_t = (\mathbf u_t, \mathbf v_t)$ is
\begin{equation*}
\begin{aligned}
\bx_{t+\delta t}\mid \bx_t
&\sim \mathcal N\!\Bigg(
\begin{pmatrix}\mathbf u_t\\ \mathbf v_t\end{pmatrix}
+\delta t
\begin{pmatrix}
h_u(\mathbf u_t,\mathbf v_t)+c\mathbf A\mathbf u_t\\
h_v(\mathbf u_t,\mathbf v_t)+\mathbf A\mathbf v_t
\end{pmatrix}, \\
&\qquad\qquad
\delta t
\begin{pmatrix}
\sigma_u^2 \mathbf I_{200} & \mathbf 0 \\
\mathbf 0 & \sigma_v^2 \mathbf I_{200}
\end{pmatrix}
\Bigg).
\end{aligned}
\end{equation*}

which is obtained by fully evolving the microscopic state over a time interval
of length $\delta t$ using the Euler--Maruyama scheme. 
Then the conditional distribution
$\hat{q}(\bx_{t+\delta t, \mathcal I} \mid \bx_t)$
obtained by restricting $\bx_{t+\delta t}$ to a local spatial patch 
$\mathcal I =\{k,\ldots,l\} \subset \{1,\ldots,200\}$ is given by the marginal of
$q(\bx_{t+\delta t} \mid \bx_t)$:
\begin{equation*}\label{eq:q_full}
\begin{aligned}
&\hat{q}(\bx_{t+\delta t,\mathcal I}\mid \bx_t) \\
&=
\mathcal N\!\Bigg(
\begin{pmatrix}
\mathbf u_{t,\mathcal I}\\
\mathbf v_{t,\mathcal I}
\end{pmatrix}
+
\delta t
\begin{pmatrix}
\big(h_u(\mathbf u_t,\mathbf v_t)\big)_{\mathcal I}
+ c\,(\mathbf A\mathbf u_t)_{\mathcal I}\\
\big(h_v(\mathbf u_t,\mathbf v_t)\big)_{\mathcal I}
+ (\mathbf A\mathbf v_t)_{\mathcal I}
\end{pmatrix},
\\
&\qquad\qquad
\delta t
\begin{pmatrix}
\sigma_u^2 \mathbf I_{|\mathcal I|} & \mathbf 0\\
\mathbf 0 & \sigma_v^2 \mathbf I_{|\mathcal I|}
\end{pmatrix}
\Bigg).
\end{aligned}
\end{equation*}
where $\mathbf A_{\mathcal I}$ denotes the submatrix of $\mathbf A$
obtained by restricting its rows to the index set $\mathcal I$.
We choose the time step $\delta t$ of the Euler--Maruyama method to be exactly
the time step used in the partial evolution scheme.

Define the one-hop neighborhood of $\mathcal I$ by
$
\mathcal I^{+}
:= \big(\mathcal I \cup \{k-1,l+1\}\big)\cap\{1,\ldots,200\}
$. For comparison, the partial evolution scheme updates only the variables within
the patch $\mathcal I$ using information from its local neighborhood $\mathcal I^{+}$:
\begin{equation*}
\begin{aligned}
\mathbf u_{t+\delta t,\mathcal I}
&=
\mathbf u_{t,\mathcal I}
+ \delta t \Big(
h_u(\mathbf u_{t,\mathcal I},\mathbf v_{t,\mathcal I})
+ c\,\mathbf A_{\mathcal I,\mathcal I^{+}}\,\mathbf u_{t,\mathcal I^{+}}
\Big) \\
&+ \sigma_u\,\delta\mathbf B_{t,\mathcal I}^u, \\
\mathbf v_{t+\delta t,\mathcal I}
&=
\mathbf v_{t,\mathcal I}
+ \delta t \Big(
h_v(\mathbf u_{t,\mathcal I},\mathbf v_{t,\mathcal I})
+ \mathbf A_{\mathcal I,\mathcal I^{+}}\,\mathbf v_{t,\mathcal I^{+}}
\Big)\\
&+ \sigma_v\,\delta\mathbf B_{t,\mathcal I}^v.
\end{aligned}
\end{equation*}
where $\mathbf u_{t,\mathcal I^{+}}$ denotes the restriction of $\mathbf u_t$ to the index set $\mathcal I^{+}$, and $\mathbf A_{\mathcal I,\mathcal I^{+}}$ is the submatrix of $\mathbf A$ obtained by restricting its rows to the index set $\mathcal I$ and its columns to $\mathcal I^{+}$.
Therefore, the conditional distribution induced by the partial evolution scheme is
\begin{equation*}
\label{eq:q_our}
\begin{aligned}
& q(\bx_{t+\delta t,\mathcal I}\mid \bx_t)
= \mathcal N\!\left(
\boldsymbol{\mu}_{t,\mathcal I},
\boldsymbol{\Sigma}_{\mathcal I}
\right), \\
& \boldsymbol{\mu}_{t,\mathcal I}
=
\begin{pmatrix}
\mathbf u_{t,\mathcal I}\\
\mathbf v_{t,\mathcal I}
\end{pmatrix}
+ \delta t
\begin{pmatrix}
h_u(\mathbf u_{t,\mathcal I},\mathbf v_{t,\mathcal I})
+ c\,\mathbf A_{\mathcal I,\mathcal I^{+}}\,\mathbf u_{t,\mathcal I^{+}}\\
h_v(\mathbf u_{t,\mathcal I},\mathbf v_{t,\mathcal I})
+ \mathbf A_{\mathcal I,\mathcal I^{+}}\,\mathbf v_{t,\mathcal I^{+}}
\end{pmatrix}, \\
& \boldsymbol{\Sigma}_{\mathcal I}
=
\delta t
\begin{pmatrix}
\sigma_u^2 \mathbf I_{|\mathcal I|} & \mathbf 0\\
\mathbf 0 & \sigma_v^2 \mathbf I_{|\mathcal I|}
\end{pmatrix}.
\end{aligned}
\end{equation*}

Comparing $\hat{q}(\bx_{t+\delta t, \mathcal{I}}\mid \bx_t)$ and $q(\bx_{t+\delta t, \mathcal{I}}\mid \bx_t)$, we note that
\begin{equation*}
\begin{aligned}
h_u(\mathbf u_{t,\mathcal I}, \mathbf v_{t,\mathcal I})
&= \big(h_u(\mathbf u_t, \mathbf v_t)\big)_{\mathcal I}, \\
h_v(\mathbf u_{t,\mathcal I}, \mathbf v_{t,\mathcal I})
&= \big(h_v(\mathbf u_t, \mathbf v_t)\big)_{\mathcal I}.
\end{aligned}
\end{equation*}
since $h_u$ and $h_v$ are applied pointwise.
Moreover, the equality
\[
(\mathbf A \mathbf u_t)_{\mathcal I}
=
\mathbf A_{\mathcal I,\mathcal I^{+}}\,\mathbf u_{t,\mathcal I^{+}}
\]
holds because the discrete Laplacian matrix $\mathbf A$ is tri-diagonal under the
Neumann boundary discretization, so that each component depends only on its
immediate neighbors.

Together, these observations imply that the conditional distribution induced by
the partial evolution scheme matches the marginal of the full evolution on the
patch $\mathcal I$, i.e.,
\[
q(\bx_{t+\delta t,\mathcal I} \mid \bx_t)
=
\hat q(\bx_{t+\delta t,\mathcal I} \mid \bx_t).
\]

Step 2: Conditional independence of $\Delta \bz_{t, \mathcal{I}}$ and $\Delta \bz_{t, \mathcal{J}}$ when $\mathcal{I} \neq \mathcal{J}$. 

From the definition of $\hat{q}(\bx_{t+\delta t,\mathcal{I}}\mid \bx_t)$, we can see that $\bx_{t+\delta t,\mathcal{I}} - \bx_t$ is independent of $\bx_{t+\delta t,\mathcal{J}} - \bx_t$ when $\mathcal{I} \neq \mathcal{J}$. Therefore, $\Delta \bz_{t, \mathcal{I}} = \bphi(\bx_{t+\delta t, \mathcal{I}}) -  \bphi(\bx_{t, \mathcal{I}})$ is conditionally independent of $\Delta \bz_{t, \mathcal{J}} = \bphi(\bx_{t+\delta t, \mathcal{J}}) -  \bphi(\bx_{t, \mathcal{J}})$. 

Step 3: $\Delta \bz^{\ast}_{t}
\approx \frac{1}{K}\sum_{\mathcal{I}} \Delta \bz^{\ast}_{t,\mathcal{I}}$

The macroscopic observables $\bz^{\ast}$ are chosen to be $(\frac{1}{200}\sum_{i=1}^{200} u(x_i, t), \frac{1}{200}\sum_{i=1}^{200} v(x_i, t))$. By definition, then the global average equals the average of the patch averages.

\end{proof}

\bibliography{main}

\end{document}